\documentclass[prb,twocolumn,superscriptaddress,showpacs,amsmath,amssymb]{revtex4-1}

\usepackage{graphicx}
\usepackage{latexsym}
\usepackage{amsmath}
\usepackage{amssymb}
\usepackage{amsfonts}
\usepackage{color}
\usepackage{xcolor}
\usepackage[normalem]{ulem}

\usepackage{bm}
\usepackage{verbatim}

\definecolor{MS-color2}{RGB}{128,0,128}

\definecolor{IB-color}{named}{green}

\usepackage{framed}
\definecolor{shadecolor}{RGB}{222,222,221}
\begin{document}

\title{Chirality selective spin interactions mediated by the moving superconducting condensate. }

 \date{\today}

\author{D. S. Rabinovich}
\affiliation{Skolkovo Institute of Science and Technology, Skolkovo 143026, Russia}
\affiliation{Moscow Institute of Physics and Technology, Dolgoprudny, 141700 Russia}

\author{I. V. Bobkova}
\affiliation{Institute of Solid State Physics, Chernogolovka, Moscow
  reg., 142432 Russia}
\affiliation{Moscow Institute of Physics and Technology, Dolgoprudny, 141700 Russia}

\author{A. M. Bobkov}
\affiliation{Institute of Solid State Physics, Chernogolovka, Moscow reg., 142432 Russia}

\author{M.A.~Silaev}
 \affiliation{Department of
Physics and Nanoscience Center, University of Jyv\"askyl\"a, P.O.
Box 35 (YFL), FI-40014 University of Jyv\"askyl\"a, Finland}


  \begin{abstract}
 We show that superconducting correlations in the presence of non-zero condensate velocity can mediate the peculiar interaction between 
   localized spins that breaks the global inversion symmetry of magnetic moments. 
  The proposed interaction mechanism  is capable of removing    fundamental degeneracies between topologically distinct magnetic textures. 
  For the generic system of three magnetic
  impurities in the current-carrying superconductor we find the  
  energy term proportional to spin chirality. 
  In  realistic superconductor/ferromagnetic/superconductor setups we reveal  
  significant  energy differences between various magnetic textures
  with opposite chiralities. 
   We calculate Josephson energies   of  junctions through   left and right-handed magnetic helices as well as through the magnetic skyrmions with  
  opposite topological charges. Relative energy shifts between otherwise degenerate magnetic textures in these setups
  are regulated by the externally controlled Josephson phase difference. The suggested low-dissipative manipulation with the skyrmion position in a racetrack geometry can be used for the advanced spintronics applications. 
  \end{abstract}

  \maketitle

 \section{introduction}
 
 Indirect interactions between localized magnetic moments mediated by conductivity electrons has been studied quite intensively since the pioneering works predicting the so-called  Ruderman-Kittel-Kasuya-Yosida
(RKKY) coupling\cite{PhysRev.96.99,doi:10.1143/PTP.16.45,PhysRev.106.893}. 
 Most of the attention has been focused on various pairwise  interactions \cite{PhysRev.115.2,PhysRev.96.99,doi:10.1143/PTP.16.45,PhysRev.106.893} between spin magnetic moments $\bm m_{1,2}$ such as the usual exchange  $E= J_{ex} (\bm m_1\cdot \bm m_2)$ 
 or the  
 Dzyaloshinskii-Moriya (DM) term \cite{PhysRev.120.91,DZYALOSHINSKY1958241, CREPIEUX1998341}
 $E= \bm D_{12} (\bm m_1 \times\bm m_2)$ which arises in system with broken inversion symmetry.
 All pairwise contributions to the interaction energy have the common property of being invariant
 with respect to the global magnetization inversion $E(\bm m) = E(-\bm m) $.
 This symmetry leads to the fundamental degeneracies between topologically distinct magnetic systems 
 which cannot be transformed into each other by the 
 global spin rotations around the symmetry axes.  The prominent example is the degeneracy between
 left-handed (upper sign) and right-handed (lower sign) magnetic spirals described by the model 
 \begin{align} \label{Eq:SpiralTexture}
 \bm m (x)= \pm \bm x \cos\alpha + \sin\alpha ( \bm y \cos\theta  + \bm z\sin\theta ) 
 \end{align}  
 with $\theta (x) = q x$ and $\alpha=const$. 
 If we assume that there is a global spin rotation symmetry 
 around $x$-axis then none of the previously known magnetic interactions
 can yield different energies of the  magnetization distributions (\ref{Eq:SpiralTexture}).
 
 Even more interesting is the setup with magnetic skyrmion\cite{Nagaosa2013,Fert2013} described by
 the spin texture 
  \begin{equation} \label{Eq:Skyrmion}
  {\bm m} = (\cos\Phi(\theta)\sin\Theta (r), \sin\Phi(\theta)\sin\Theta (r), \cos\Theta(r) )
  \end{equation}
  where $(\theta,r)$ are the polar coordinates. 
 The azimuthal structure corresponds to the magnetic vortex given by $\Phi(\theta) = \kappa_v \theta + \nu\pi/2$, 
  where the integer $\kappa_v$ is vorticity and $\nu =\pm 1$ is the helicity determined by the sign of Dzyaloshinskii – Moriya interaction 
  \cite{Nagaosa2013}.
  The spin at the core points up (down)  while at the perimeter it tends to rotate 
  to the opposite direction. The states with $[\cos\Theta (r)]_0^\infty = \pm 2$ correspond to different polarity\cite{han2017skyrmions}.  These two options lead to 
  the different topological charges
  $Q=  \pm \kappa_v$ characterizing two energetically degenerate magnetic states. 
  The sign of topological charge
   determines the flux of emergent magnetic field and thus
  the sign of topological Hall resistivity measured in experiments \cite{Neubauer2009,Schulz2012,Liang2015}.
 From the general definition
  of topological charge
  $Q = \frac{1}{4\pi} \int d^2\bm r ~\bm m \cdot (\partial_x \bm m \times \partial_y \bm m)$ one can see  that
 in  the absence of external magnetic field none of  previously  known spin interactions can remove the 
  degeneracy with respect to $Q \to -Q$ due to the magnetization inversion $\bm m\to -\bm m$. 
   The proposed chirality-selective interaction
  will be shown to fix the  ground state value of polarity and
  therefore $Q$ thus providing in principle the field-independent 
  contribution to the topological Hall effect. 
  
  In this paper we point out the fundamental spin interaction which removes the above-mentioned
  degeneracies between magnetic textures. This contribution appears in 
  the presence of moving superconducting condensate, or in other words the current-carrying superconducting correlations. 
  The possibility of such interaction can be understood from the symmetry arguments.  
  Let us consider the generic example of three magnetic moments $\bm m_{1,2,3}$
  localized at spatially separated points in the  metal which does not contain other magnetic moments and in the absence of external magnetic field. 
 The energy proportional to spin chirality $E_{ch}= E_a \bm m_1 \cdot(\bm m_2\times \bm m_3) $
 is possible only if the scalar prefactor $E_{a}$ changes the sign under the time-reversal transformation $\cal T$. 
 Since we assume that there are no other magnetic moments in the host metal, such scalar $E_a\neq 0$ cannot be constructed in the normal state. In the next Section we demonstrate this by the explicit calculation. 
 However the
 $\cal T$-odd scalar exists in superconducting state where the condensate moves with non-zero velocity $\bm v_s\neq 0$. This state breaks the time reversal symmetry and one can choose $E_a$ to be the projection of  
 superfluid velocity on the some anisotropy axis determined e.g. by the spatial configuration of magnetic impurities.  
  Thus in superconducting states with $\bm v_s\neq 0$ chirality-selective triple spin interactions are generically possible
  \footnote{ The chirality-sensitive terms in the free energy were calculated for the system consisting of the Josephson junction through  magnetic trilayer
 \cite{Kulagina2014}. It has been obtained that the presence of scattering barriers separating ferromagnetic regions is crucial for such terms to be non-zero. In the present work we show that the chirality-selective energy arise in the generic problem with three magnetic impurities and no extra conditions are needed. Also we demonstrate that such energy contributions appear in the systems with continuous spin textures like magnetic spiral and skyrmion.  
}.
  In addition to the projection of $\bm v_s$ the 
 amplitude of $E_a$ contains prefactor determined by the distance between impurities as demonstrated in this paper.
  {In principle one can expect that even in the  normal magnetic system the spin-transfer torques mediated by resistive currents can depend on the spin chirality. However, this is a non-equilibrium effect which is beyond the scope of the present paper. }

 The mechanism discussed above can be very important
 for different hybrid ferromagnet/superconductor (FM/SC) structures as well as for the interacting magnetic impurities in superconductors 
 \cite{L.U.H.1965,Shiba1968,Rusinov1969,PhysRevLett.113.087202,PhysRevB.84.224517,RevModPhys.78.373} and 
  magnetic adatoms placed on top of the superconducting surface
  \cite{PhysRevB.85.144505, PhysRevB.85.020503, PhysRevB.84.195442, PhysRevB.89.115109, PhysRevLett.111.206802, 
  PhysRevLett.111.147202, PhysRevLett.111.186805, PhysRevB.88.180503, PhysRevB.88.020407, 
  PhysRevB.89.180505,PhysRevB.88.155420,Yazdani1767}. 
  Such systems are in the focus of attention nowadays in connection with topological quantum computations and 
  advanced spintronics applications based on the low-dissipative manipulations with magnetic textures.

  The paper is organized as follows. In Sec.~\ref{Sec:generic} we consider the generic mechanism of triple spin interactions by the example of three magnetic moments located at spatially separated points. In Sec.~\ref{Sec:josephson} the contribution of triple spin interactions to the Josephson energy of junctions via magnetic helices and skyrmions is calculated. Sec.~\ref{Sec:discussion} is devoted to the discussion of the results and their possible applications.

 \section{ Generic example of triple spin interactions}
  \label{Sec:generic}
\begin{figure}[!tbh] 
  \centerline{\includegraphics[clip=true,width=3.4in]{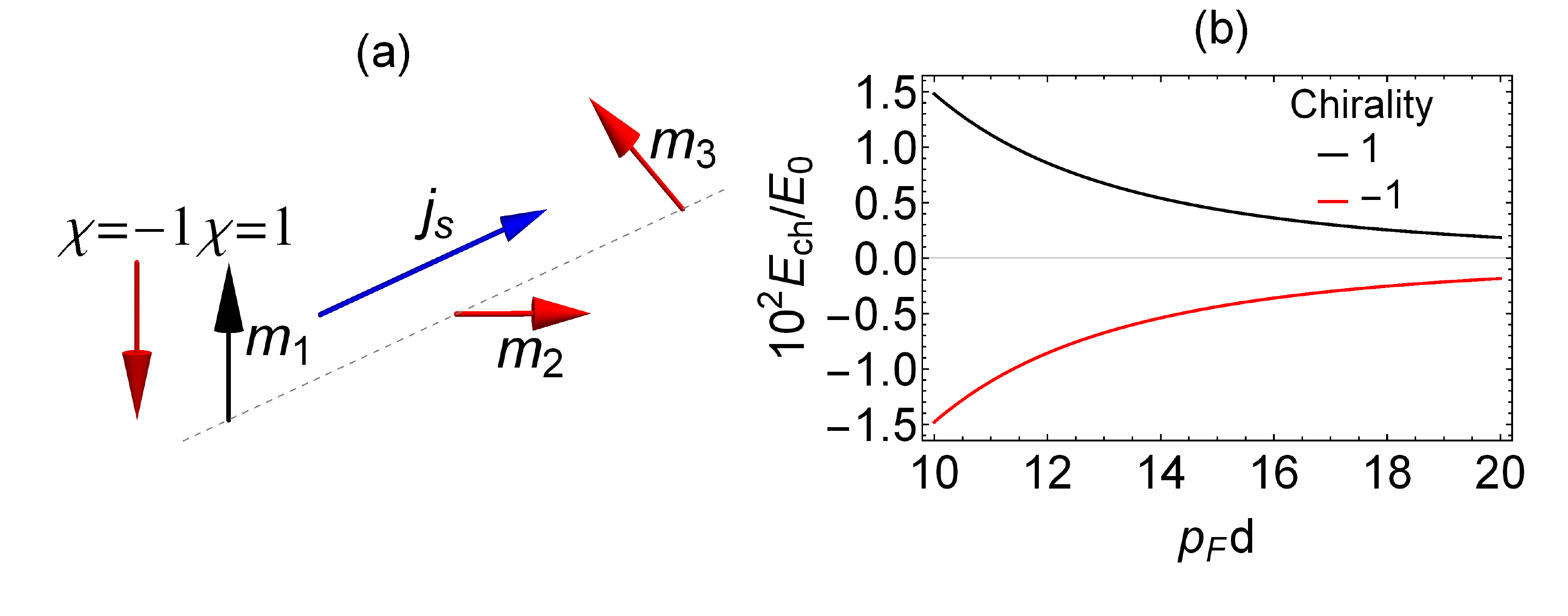}}
  \caption{(a) System of three magnetic impurities. 
  (b) Chirality-sensitive contribution into the 
  interaction   energy, $E_0=p_Fv_s(N_0 J)^3.$}
  \label{Fig:TripleSpin}
 \end{figure}

 Let us assume that three magnetic impurities with moments $\bm m_{1,2,3}$ residing at the points 
  $\bm r_l =(x_l, 0, 0)$ with $x_1=0$, $x_2=d$, $x_3=2d$ along the $x$-axis (Fig.\ref{Fig:TripleSpin}a).
  They are described by Hamiltonian $J \delta(\bm r - \bm r_l)  (\bm \sigma \bm m_l)$ giving rise to 
  the following contribution to the free energy of the system  $E= J \sum_l \bm m_l\cdot \bm s(\bm r_l)$ where $\bm s(\bm r_l) = (T/4) \sum_\omega {\rm Tr}[{\bm \sigma} \hat G(\omega, \bm r_l,\bm r_l)]$ is the average spin density of conducting electrons at the point $\bm r_l$ expressed through the Matsubara Green function $\hat G(\omega, \bm r, \bm r^\prime)$ which in general depends in two coordinates $\bm r$ and $\bm r^\prime$. The non-zero contribution to $E$ containing triple product of $\bm m_{1,2,3}$ is provided by the second-order correction to the GF $\hat G^{(2)}(\omega, \bm r, \bm r^\prime)= J^2 \sum_{k\neq j} 
  (\bm \sigma \bm m_k) (\bm \sigma \bm m_j)
  \hat G^{(0)}(\omega, \bm r-\bm r_k) 
   \hat G^{(0)}(\omega, \bm r_k-\bm r_j) 
   \hat G^{(0)}(\omega, \bm r_j-\bm r^\prime)$,
   where $\hat G^{(0)}(\omega, \bm r) $ is the GF of the superconductor without magnetic impurities taking into account the modification of spectrum due to the condensate velocity. 
   Then the  unusual contribution into the interaction energy which involves  three magnetic moments takes the form 
  \begin{align} \label{Eq:3momentEnergy0}
   & E_{ch} = \frac{i J^3}{4} T \sum_\omega 
  \sum_{l\neq k\neq j} \bm m_l\cdot (\bm m_k \times \bm m_j) 
   \\ \nonumber
  & {\rm Tr} 
  \left[  \hat G^{(0)}(\omega,\bm r_l - \bm r_k) 
  \hat G^{(0)}(\omega,\bm r_k - \bm r_j) 
  \hat G^{(0)}(\omega,\bm r_j - \bm r_l) \right]
  \end{align}    
  This equation can be transformed as 
  {
  \begin{align} \label{Eq:3momentEnergy0_1}
  & E_{ch} = \bm m_1\cdot (\bm m_2 \times \bm m_3) \frac{i J^3}{4}
  T \sum_\omega  {\rm Tr} [ 
  \\ \nonumber
   &    \hat G^{(0)}(12) 
  \hat G^{(0)}(23) 
  \hat G^{(0)}(31) 
  -
   \hat G^{(0)}(13) 
  \hat G^{(0)}(32) 
  \hat G^{(0)}(21) 
  +
  \\ \nonumber
   &\hat G^{(0)}(23) 
  \hat G^{(0)}(31) 
  \hat G^{(0)}(12) 
  -
   \hat G^{(0)}(21) 
  \hat G^{(0)}(13) 
  \hat G^{(0)}(32) 
  +
  \\ \nonumber
   &\hat G^{(0)}(31) 
  \hat G^{(0)}(12) 
  \hat G^{(0)}(23)
  -
   \hat G^{(0)}(32) 
  \hat G^{(0)}(21) 
  \hat G^{(0)}(13) ]
    \end{align}
    }
  where we denote $\hat G^{(0)}(kj) = \hat G^{(0)}(\omega, \bm r_k - \bm r_j) $. 
   This expression can be simplified as follows
  \begin{align} \label{Eq:3momentEnergy}
  & E_{ch}(d) = \frac{3i J^3 \chi }{4} T
  \\ \nonumber
  & \sum_\omega {\rm Tr} 
   [ \hat G^{(0)}(\omega,-d)  \hat G^{(0)}(\omega,-d) \hat G^{(0)}(\omega,2d) - 
  \\ \nonumber
  & \hat G^{(0)}(\omega,d)  \hat G^{(0)}(\omega,d) \hat G^{(0)}(\omega,-2d) ]
  \end{align}    

  where $\chi = \bm m_1\cdot (\bm m_2\times \bm m_3)$ is the 
  scalar spin chirality and $\hat G(\omega,x)$ are the Matsubara Green's functions (GF) taken along the $x$ axis at $y=z=0$.
 The non-zero triple interaction $ E_{ch}\neq 0$ appears in the presence of condensate velocity which we assume to be directed 
 as $\bm v_s = v_s \bm x$. In the momentum-space representation 
 of GF the frequency acquires Doppler shift 
 $\tilde \omega =  \omega + i  (\bm p_F  \bm v_s) $,
 where $\bm p_F$ is the Fermi momentum. We expand the GF to the first order by the 
 Doppler shift $\hat G^{(0)}=G^{(0)}_0 + G^{(0)}_1$. 
 In momentum space the GF at $v_s=0$ is given by $G^{(0)}_0 = (\xi_p \tau_3 - i\omega\tau_0 + \Delta\tau_1)/(\xi_p^2 + \omega^2 + \Delta^2 )$ 
 where $\xi_p = v_F(p-p_F) $, 
    and $v_F$
 is the Fermi velocity. The correction to the first-order in condensate velocity is $G^{(0)}_1= i(\bm p_F \bm v_s) dG^{(0)}_0/d\omega$.
 Next, we calculate the real-space representation GF
 \begin{align} \label{Eq:GF0_real_space}
 & \hat G^{(0)}_0 (x) = 
 \\ \nonumber
 &\pi N_0\frac{e^{-\Omega|x|/v_F}}{p_F|x|}  \left[ 
 \tau_3 \cos(p_Fx)  + i \tau_3 \hat g \sin(p_F|x|) \right]
 \\ \label{Eq:GF1_real_space}
 & \hat G^{(0)}_1 (x) =  \frac{i\pi N_0}{p_F x} (v_s p_F) e^{-  \Omega|x|/v_F} \cos (p_Fx) 
 \tau_3  \frac{d\hat g}{d\omega}
 \end{align}  
 where
 $N_0$ is the Fermi-level density of states, 
 $\hat g = (\omega \tau_3+ \Delta\tau_1)/\Omega$, 
 where $\Omega = \sqrt{\omega^2 +\Delta^2}$.
 Substituting expressions (\ref{Eq:GF0_real_space}), 
 (\ref{Eq:GF1_real_space}) to Eq.(\ref{Eq:3momentEnergy}) we obtain the triple energy as a function of the distance between localized spins.

    First of all let us discuss the general result for the triple interaction which is obtained from Eq.(\ref{Eq:3momentEnergy})
  in the first order by the condensate velocity 
  \begin{align} \label{Eq:3momentEnergy1_1}
   E_{ch}(d) =  3\chi p_Fv_s \Delta^2 \frac{(\pi N_0 J)^3}{X^3}   
   T \sum_\omega  \frac{ e^{-4 d \Omega/v_F} }{\Omega^3} 
  \end{align}   
  where we denote $X=p_Fd$. 
  From Eq.(\ref{Eq:3momentEnergy1_1}) it is clear that 
  the triple energy vanishes in the normal state $\Delta=0$
  as it should be according to the general symmetry consideration in the Introduction. 
  
  Now let us consider the limiting case of small temperatures 
  $T\ll \Delta$ and distances $\Delta d/v_F \ll 1$. Then we can integrate Eq.(\ref{Eq:3momentEnergy1_1}) over the Matsubara frequencies to get 
  \begin{align} \label{Eq:3momentEnergy2}
   E_{ch}(d) =  3\chi p_Fv_s \frac{\pi^2 (N_0 J)^3}{2X^3}   
   \end{align}   
 Note that remarkably we get the result which does not contain Friedel oscillations with the scale of the Fermi wavelength. This shows that the triple spin interaction  is mediated exclusively by the Cooper pairs 
 without the participation of single-particle excitations. 
 
This dependence is shown in  Fig.\ref{Fig:TripleSpin}b with red and black  curves corresponding to  the opposite chiralities $\chi=\pm 1$. 
This interaction is much smaller than the usual RKKY exchange which has the amplitude of the order of $E_{ex} \approx E_F (N_0 J)^2/(p_F d)^2 $, where $E_F$ is the Fermi energy. Therefore $E_{ch}/E_{ex} \sim N_0 J (v_s/v_F)  (1/p_F d)$.
This ratio contains three small parameters because the  perturbation theory that we used is valid when $N_0 J  \ll 1$. In addition, the condensate velocity is always much smaller than the Fermi velocity $v_s\ll v_F$ and the distance between spins is larger than the Fermi wavelength $p_Fd>1$. 
However, in contrast to the usual exchange interaction the energy $E_{ch}$ breaks the symmetry with respect to $\chi=\pm 1$ so that even if its amplitude is small it can provide the new way of controlling magnetic structures such as magnetic helices and skyrmions discussed below.

 \section{Anomalous Josephson energy}
 \label{Sec:josephson}
  \subsection{Model}
 This interaction mechanism (\ref{Eq:3momentEnergy1_1},\ref{Eq:3momentEnergy2}) is fundamentally different from usual exchange and DM interactions. It can show up in various systems hosting magnetic moments and the superconducting condensate with non-zero velocity. The important subclass of such systems are the Josephson junction with spin-textured interlayers. The analog of condensate velocity in Josephson systems is the phase difference between superconducting electrodes. Thus for the fixed phase difference one can expect the energy shift between magnetic structures with opposite chiralities. {In particular, the discussed chiral contribution leads to the fact that the dependence of the free energy on the phase difference becomes asymmetric, as it was demonstrated in Ref.~\onlinecite{Kulagina2014}. } 
 Examples that we consider here include left- and right-handed magnetic spirals (\ref{Eq:SpiralTexture}) and magnetic skyrmions with opposite topological charges (\ref{Eq:Skyrmion}). 

In case of the weak proximity effect with large interface barrier the Josephson current-phase relation (CPR) can be expressed as
 \begin{align}
j=j_o \sin \varphi + j_a \cos \varphi.
\label{current_harmonics}
\end{align}
Here the first term is the ordinary 
contribution with the amplitude $j_o$. The second term with anomalous phase shift 
is in general proportional to the spin chirality $j_a\propto \chi$ which can be introduced 
 in various different ways depending on the particular system under consideration.
 
 
{
The anomalous Josephson effect (\ref{current_harmonics}) with 
$j_a\neq 0$ can be considered as the inverse magnetoelectric effect. In superconducting 
systems magnetoelectric effects are especially interesting, because they can manifest 
 itself in many different ways. Responding to the applied exchange field   superconducting systems with spin-orbit coupling (SOC)   can generate a spontaneous current \cite{Bobkova2004,Dolcini2015,Pershoguba2015,Malshukov2016,Mironov2017}, or experience a transition to the phase modulated helical state 
 \cite{edelshtein1989,Barzykin2002,PhysRevB.70.104521,PhysRevLett.94.137002,PhysRevB.76.014522,PhysRevB.92.014509}.
  The particular type of the response depends on the geometry of the system.
 The  anomalous Josephson effect is a manifestation of the magnetoelectric effect, specific for Josephson junctions. It was proposed for SOC interlayers under the applied magnetic field and  for Josephson junctions with  noncoplanar magnetic interlayers \cite{krive04, Braude2007, Asano2007,Reynoso2008,Eschrig2008, PhysRevLett.101.107005,Tanaka2009,Grein2009, Zazunov2009, Liu2010,Malshukov2010,Alidoust2016,Brunetti2013,Yokoyama2014,
Kulagina2014,Moor2015,Moor2015a,Bergeret2015,Campagnano2015,Mironov2015,PhysRevB.92.125443,
Kuzmanovski2016,Zyuzin2016,Silaev2017,Bobkova2017}
  or under the nonequilibrium quasiparticle injection \cite{PhysRevB.94.134506}.
   This effect has been recently observed in the Josephson junctions with spin-orbital interaction \cite{Szombati2016, 1806.01406,Murani2017}. The signatures of anomalous shift have been seen in the trilayer ferromagnetic Josephson structure\cite{Glickeaat9457}.
   The interpretation of the 
   anomalous phase shift in terms of the inverse magneto-electric effect was proposed in \cite{PhysRevB.92.125443}.
  

   The anomalous current term in (\ref{current_harmonics}) can be rewritten in the form $j=j_c \sin (\varphi - \varphi_0)$ with $j_c = \sqrt{j_o^2 + j_a^2}$ and $\tan \varphi_0 = -j_a/j_o$. This leads  to appearance of the anomalous contribution to the Josephson energy $\frac{2e E_J}{\hbar} = j_c[1-\cos(\varphi - \varphi_0)]$:
 \begin{align}
\frac{2e E_J}{\hbar}= \sqrt{j_o^2+j_a^2}- j_o \cos \varphi + j_a \sin \varphi.
 \label{energy_harmonics}
 \end{align}
  The last term here  provides the anomalous energy contribution and it has different signs 
 for the spin textures of different chirality. 
  }%
 Below we calculate the amplitudes $j_o$, $j_a$ using the powerful machinery of 
 quasiclassical Usadel theory\cite{Bergeret2005a}, which works in weak ferromagnets when the exchange splitting 
 is much smaller than the Fermi energy. This condition is satisfied 
 for the transition-metal compounds of the MnSi family, where the exchange field can be estimated as 
 $h \sim 100 $ meV being is much less than the Fermi energy $\sim 1$eV
 \cite{Lee2007,Jonietz2010} 
 
  Previously, no  anomalous Josephson effect has been found in a number of works which considered
   non-coplanar structures, such as magnetic spiral \cite{Volkov2006}, vortex\cite{Kalenkov2011} and skyrmion\cite{Yokoyama2015} in "weak FMs",  that is described within the quasiclassical Usadel theory\cite{Buzdin2005,bergeret2005odd}.   
 The general reason for that as  identified recently \cite{Silaev2017} is the artificial  symmetry 
 $j(\bm m)=j(-\bm m)$ which appears in quasiclassical equations. Together with the time-reversal symmetry $j(\bm m, \varphi) = -j(-\bm m, -\varphi)$ it leads to the symmetry of the Josephson current $j(\varphi) = -j(-\varphi)$, which prohibits the anomalous current. 
 In order to get rid of this symmetry  we assume the presence of interface spin-filtering barriers 
 characterized by the dimensionless polarization vector ${\bm P}$ . Such barriers can be
  described by the effective boundary conditions \cite{bergeret2012electronic,Bergeret2012a,Eschrig2015}.
 In this case the situation $j(\bm m, \bm P) \neq j(-\bm m,-\bm P)$ is possible. This trick allows for breaking the symmetry of Josephson current $j(\varphi) \neq -j(-\varphi)$ and, consequently, for the realizations of the anomalous Josephson effect. 
 Simultaneously this allows for extending the range of suitable materials which can be found in the B20 family of itinerant cubic helimagnets, MnSi, (Fe,Co)Si, and FeGe  
 \cite{Ishikawa1977, Pfleiderer2001,Uchida2006}.
 
 
 

 
 To analyze a proximity effect in FS system we considered
linearized Usadel equation for the quasiclassical anomalous function which takes
into account triplet and singlet superconducting
correlations\cite{bergeret2005odd}. We presented anomalous function in the
ferromagnetic region in the form
 $\hat{f}=f_0\hat{\sigma}_0+f_x\hat{\sigma}_x+f_y\hat{\sigma}_y+f_z\hat{\sigma}_z$.
In this expansion the first term corresponds to the singlet
component and the last three terms correspond to the triplet
components. As we focus here on the equilibrium problem 
 we work in the Matsubara frequency $\omega$ representation. The equations for
coefficients $f_i$ are (for $\omega>0$):
 \begin{align}\label{Eq:Usadel}
 & (D\nabla^2_x-2\omega) f_0 - 2i {\bm h}\cdot {\bm f}=0,\\
 & (D\nabla^2_x - 2\omega) {\bm f} - 2i   f_0 {\bm h} =0 \nonumber
 \end{align}
 where ${\bm f}=(f_x,f_y,f_z)$.
 %
 Spin-filtering barriers at SC/FM interfaces are  described by the generalized  Kuprianov-Lukichev boundary conditions \cite{KL}, that include  spin-polarized 
 tunnelling at the SF interfaces \cite{BergeretVerso2012, Machon2013, Machon2014, EschrigBC2015} 
 \begin{align}\label{bc}
 & (\gamma \partial_n + G_S) {\bm f}  = - i G_S (\bm P\times \bm f) 
 \\
 & (\gamma \partial_n + G_S) f_0 = \sqrt{1-P^2} F_{S} \nonumber
 \end{align}
 where $F_{S}=\Delta/\sqrt{|\Delta|^2+\omega^2}$ and $G_S=\omega F_S/\Delta$ are 
 the anomalous and normal GF in the superconducting region,
 $\partial_n = \mp \nabla_x$ corresponds to the left and right interfaces, at $x=0$ and $x=d$ respectively. 
 Here $D$ is diffusion coefficient, ${\bm h}\parallel \bm m$  is the exchange
 field parallel to the magnetization direction, $\omega$ is the
 Matsubara frequency.
 
 The current is given by 
 \begin{align} \label{Eq:Current1}
 \frac{e\bm j}{2\pi T\sigma_n} = \sum_{\omega>0} {\rm Im} (f_0^*\nabla f_0 - f_t^*\nabla f_t)
 \end{align}


   %
   The simplest example of SC/FM/SC system which supports anomalous Josephson effect consists of the spin-filter barrier with polarization $\bm P$ and 
 two weak FMs with misaligned magnetizations  $\bm m_{1,2}$. It has been shown \cite{Silaev2017,PhysRevB.96.064519} 
 that the anomalous current in such system looks like $j_{an} = \chi I_{an} \cos\varphi$, where the chirality is
 $\chi= \bm P\cdot (\bm m_1\times \bm m_2)$. 
 {
 The corresponding term in Josephson energy  reads
 \begin{equation} \label{Eq:AnomalousEnergy}
  E_{ch} = \frac{\hbar}{2e} I_{an} \sin\varphi \bm P\cdot(\bm  
  m_1\times \bm m_2)
  \end{equation}
  }
 For the simplest trilayer structure  under the conditions of the fixed phase difference  
  the energy (\ref{Eq:AnomalousEnergy}) fixes the sign of $\chi$ in the ground state. 
  
  Introducing the spin filtering barrier with a polarization $\bm P$ is not the only way to violate the symmetry $j(-\bm m) = j(\bm m)$ in a trilayer Josephson setup and, consequently, to have the anomalous Josephson current. There are a number of papers, where the anomalous contribution to the Josephson current in S/F/S junctions with a trilayer magnetic interlayer has been obtained in the framework of other models \cite{Braude2007, Asano2007,Eschrig2008,Grein2009,Kulagina2014,Liu2010,Mironov2015,Silaev2017}. Here we reproduce this known result in the framework of the model with two weak FMs with misaligned magnetizations and a spin-filtering barrier just because it is in line with our main consideration of spin-textured ferromagnetic interlayers presented below.
  
  In setups which are more complicated than the model layered Josephson structure
  the expression for anomalous energy can be more involved, but still it has the same general feature 
  of being odd in the magnetic momentum and containing the superconducting phase difference of the condensate velocity 
  to restore the time invariance. Such unusual energy contributions can lead to the interesting effects removing the 
  degeneracy by energy between otherwise degenerate spin textures. 
  
 \subsection{Magnetic helix}
  
\begin{figure*}[htb!]
 \centerline{$
 \begin{array}{c}
 \includegraphics[width=0.4\linewidth]{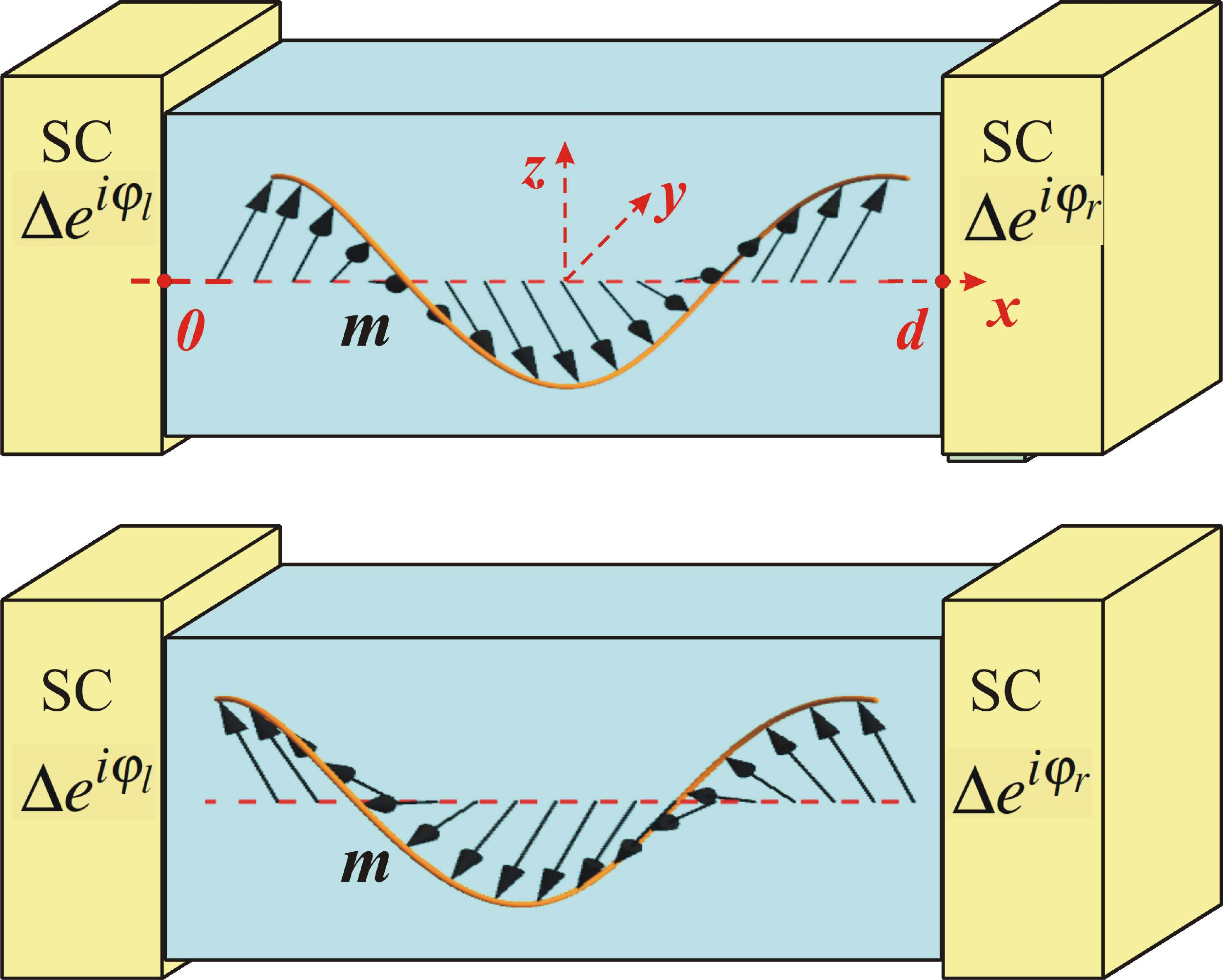}
 \put (-120,163) { \large(a) $\chi_1>0$ }
 \put (-120,75) { \large(b) $\chi_1<0$ }
 \;\;\;\;\;\; \includegraphics[width=0.45\linewidth]{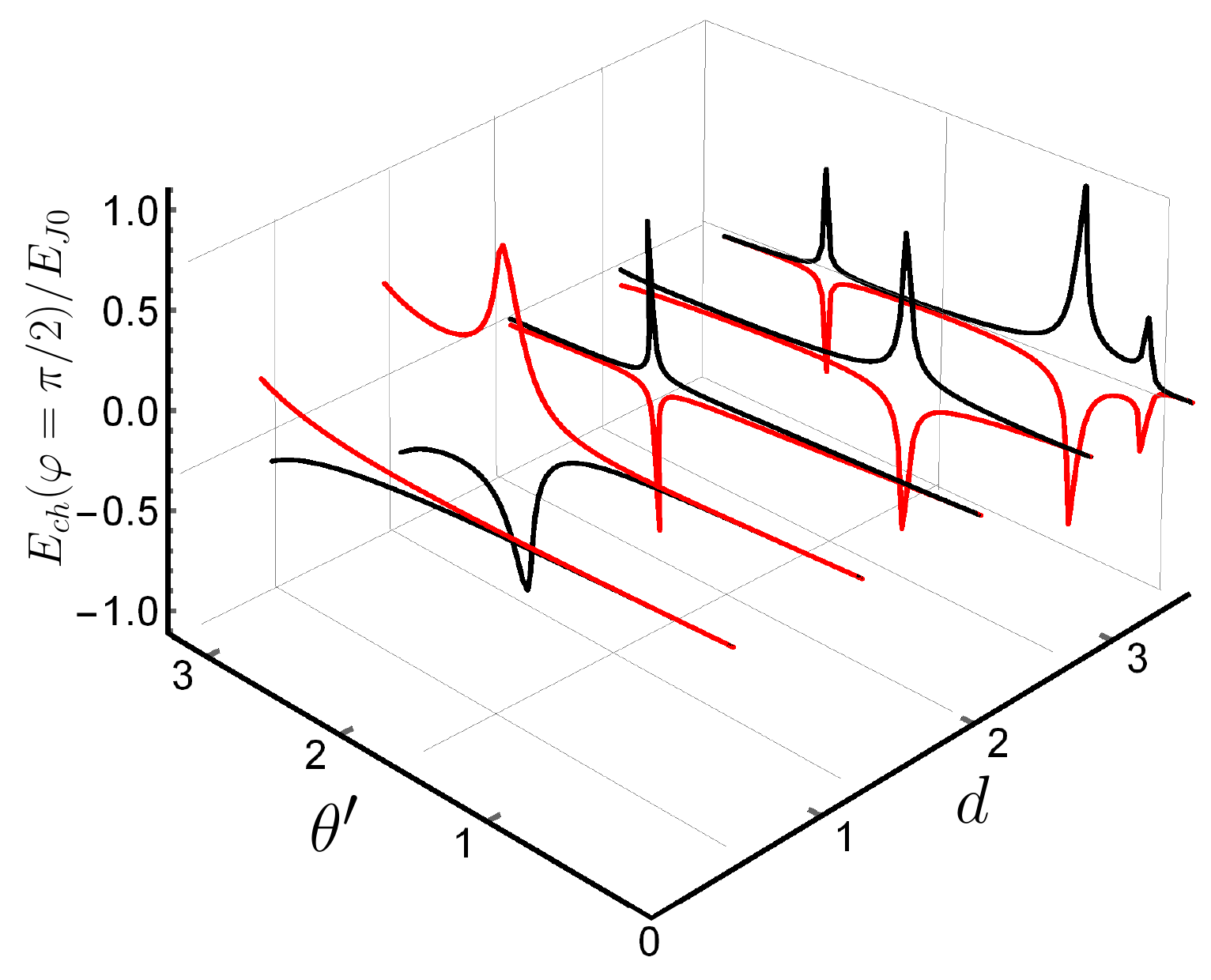}
 \put (-190,160) { \large (c) }
  \end{array}$}
 \caption{\label{Fig:energy_helix} (Color online) 
 (a,b) Josephson junctions through the left- and right-handed magnetic helices characterized by the chiralities 
 $\chi_1>0$ and $\chi_1<0$, respectively. The interface barrier polarizations $\bm P$ is assumed to be aligned with the local magnetization 
 $\bm P_l\parallel \bm m(x=0)$ and $\bm P_r\parallel \bm m(x=d)$.
 (c) Chiral contribution $E_{ch}(\varphi = \pi/2)/E_{J0}$ to the Josephson energy for the junction via magnetic helix.
  Black and red lines correspond to different chiralities ${\rm sgn}\chi_1 = \pm 1$. 
  The numerical parameters are $P=0.5$, $\alpha=\pi/4$, $\gamma = 1$, $h=40$.
 }
 \end{figure*}

  Let us consider the example of the helical magnetic configuration described by the pattern
  (\ref{Eq:SpiralTexture}). 
   The sketch of the system is presented in Fig.~\ref{Fig:energy_helix} a,b.
 In addition, we assume that there are spin-filtering barriers described by the polarizations $\bm P_l$ and 
 $\bm P_r$ at the left ($x=0$) and right ($x=d$)
 FM/SC interfaces, respectively. 
  We demonstrate that the chiral spin interaction given by the last term in Eq.(\ref{energy_harmonics})
 selects the particular 
 chirality of magnetic configuration, that is the sign of the first term 
 in (\ref{Eq:SpiralTexture}). Given that the system has an additional global spin rotation symmetry
 this is equivalent to the change in the sign of $\theta$ or the 
  swirling direction of the magnetization determined by the sign of azimuthal angle 
 gradient $\theta^\prime \equiv \nabla_x \theta$.
 
  \subsubsection{Analytical consideration}
 
 %
Generally the homogeneous Eqs.(\ref{Eq:Usadel}) have the solutions
of two types which are the short-range and long-range modes with
the scales $\xi_h=\sqrt{D/2h}$ and $\xi_\omega=\sqrt{D/2|\omega|}$
correspondingly. Hereafter we assume that $\xi_h$ is the smallest
length of the problem such that the spatial dependencies of
exchange field and geometrical factors are characterized by the
scales $\gg \xi_h$. 
Under such conditions we search for the short-range solutions of Eqs.(\ref{Eq:Usadel}) in the form 
$\bm f_t = \bm m f_{sr} $, where $\bm m$ is the local direction of magnetization/exchange field, and at the 
left interface we find
  \begin{align} \label{Eq:SRSC0}
 & f_0 (x) =  X_1 e^{- \lambda x} + X_2 e^{- \lambda^* x} 
 \\ \label{Eq:SRTC0}
 & f_{sr} (x) = X_1 e^{- \lambda x}  - X_2 e^{- \lambda^* x} ,
 \end{align}
 where $\lambda = k_h e^{i\pi/4}$, $k_h=\xi_h^{-1}$ and the coefficients $X_{1,2}$ determined by the boundary conditions.

 The structure of long-range modes cannot be determined analytically for the general magnetization pattern. 
 Here we consider the particular case of magnetic helix (\ref{Eq:SpiralTexture}).
 The two long-range modes are given by the superposition
 \begin{align}
  \bm f_{lr} = f_{\theta}\bm n_\theta + f_{\alpha}\bm n_\alpha 
 \end{align}
  in terms of the two orthogonal vectors 
 \begin{align} \label{Eq:n_theta}
 & \bm n_\theta =- \partial_\theta \bm m
 \\
 \label{Eq:n_alpha}
& \bm n_\alpha =-   
(\partial_x \bm n_\theta\cdot \partial_\alpha \bm m)
\partial_\alpha \bm m
\end{align}
  which are also orthogonal to $\bm m$. 
  Note the physical reason which explains the existence of two long-range modes is the noncoplanar magnetic texture which generates two independent vector fields $\bm n_\alpha$ and 
  $\bm n_\theta$ orthogonal to the magnetic texture.
 {  From these vectors $\bm n_\alpha$, $\bm n_\theta$ in combination with the normalized exchange field $\bm m$
 and spin filter polarization $\bm P$ one can combine three different spin chiralities 
  \begin{align} \label{Eq:chi1}
  & \chi_1= \bm P\cdot (\bm n_\alpha\times \bm n_\theta) 
  \\ \label{Eq:chi2}
  & \chi_2 = \bm P\cdot (\bm m \times \bm n_\theta)
  \\ \label{Eq:chi3}
  & \chi_3 = \bm P\cdot (\bm m \times \bm n_\alpha)
  \end{align}
  Note that 
    $\chi_1$ is qualitatively different from $\chi_{2,3}$. 
     While $\chi_{2,3}$ require a misalignment between the local exchange field and the interface polarization,  $\chi_1\neq 0$ even if $\bm P\parallel \bm m$ at the barrier. Therefore $\chi_{2,3}$ are associated with the external "interface" chirality of the structure, and $\chi_1 = (\bm P \bm m)\chi_{in}$, where $\chi_{in} = \bm m\cdot (\bm n_\alpha \times \bm n_\theta)$ is the internal chirality of the magnetic texture. Consequently, $\chi_1$ is the quantity that determines the anomalous Josephson effect in case when both the spin-filtering polarization and spin rotation come from the same exchange field.
  
 %
 

 The CPR can be calculated analytically in long junction $d>\xi_\omega$ and 
 "slow" magnetic helix $\theta^\prime \ll k_\omega$. 
 It means that the helix magnetization rotates slowly on the scale $\xi_N$.
 %
   %
%
%
 %
The resulting CPR takes the form of Eq.~(\ref{current_harmonics}) with anomalous $j_a$ current contribution
given by the superposition of three parts $j_a = j_{ex} + j_{in} + j_{mix}$, where
the three contributions are determined by the different chiralities 
\begin{align} \label{Eq:j_in}
 & j_{in} \propto (\chi_{1l}+\chi_{1r})
\\ \label{Eq:j_ex}
 & j_{ex}\propto (\bm P_l \bm m_{l}+\bm P_r \bm m_{r})
 \frac{( \chi_{2l}\chi_{3r}-\chi_{3l}\chi_{2r})}{\theta^\prime \cos \alpha}
 \\ \label{Eq:j_mix}
 & j_{mix} \propto (\chi_{1l}+\chi_{1r})
 \left[\chi_{2l}\chi_{2r}+\frac{\chi_{3l}\chi_{3r}}{\theta^{\prime 2} \cos^2 \! \alpha }\right]
\end{align}
 %
 where
 $\chi_{i l}$ and $\chi_{i r}$ are the values of chiralities (\ref{Eq:chi1})-(\ref{Eq:chi3}) evaluated at left and right FM/SC interfaces
 and   $\bm m_{l} = \bm m (0)$ and $\bm m_{r} = \bm m (d)$. 
 The amplitudes of current contributions are given in Appendix. 
 To understand the physical meaning of all three terms we consider different physical situations. 

{\it Homogeneous ferromagnet. Interface chirality.} At first we consider the limit of homogeneous ferromagnet $\theta^\prime = 0$. In this case $\chi_1 = 0$, because it is entirely determined by the "internal chirality" of the ferromagnet texture. But $\chi_2$ and $(\chi_3/\theta^\prime \cos \! \alpha)$ can be nonzero due to noncoplanarity of $\bm m$ and the interface polarizations $\bm P_l$ and $\bm P_r$. 
It can be easily deduced from Eqs.~(\ref{Eq:j_ex}),(\ref{Eq:j_mix}) that in this case the only nonzero contribution to the anomalous current is $j_{a} \sim (\bm P_l \bm m + \bm P_r \bm m)[\bm m\cdot (\bm P_l \times \bm P_r)]$. Therefore, the anomalous current is not only proportional to the mutual  chirality $\bm m\cdot (\bm P_l \times \bm P_r)$ of the three characteristic magnetic vectors existing in the system. The scalar product $\bm P_{l,r} \bm m$ also must be nonzero. This result has already been obtained in Ref.~\onlinecite{Silaev2017}. The reason is that for a weak ferromagnet with $h \ll \varepsilon_F$ the "interface chirality" factor $\bm m\cdot (\bm P_l \times \bm P_r)$ by itself does not satisfy the symmetry $j(\bm m)=j(-\bm m)$, which appears in the quasiclassical equations. 

 {\it General case.}
 When all chiralities are non-zero $\chi_{1,2,3} \neq 0$ all three terms contribute to the anomalous current.
 However, in the limit $\xi_\omega\theta^\prime \ll 1$ we can neglect $j_{in}$ so that $j_a = j_{ex}+j_{mix}$. 
 In general $j_{ex}$ is an even function of $\theta^\prime$, therefore it does not depend on the internal chirality of the texture and is controlled by the interface chirality. On the contrary, $j_{mix}$ has different signs for the opposite helix chiralities. The main role of nonzero $\chi_2$ and $\chi_3$ here is to generate long-range spin-triplet pairs due to noncollinearity of the internal and boundary magnetizations at the interfaces.

 {\it Internal chirality.}
 Further we focus on the case  when $\theta^\prime \neq 0$ 
but the interface polarization is aligned with the local direction of the magnetization at each of the interfaces 
$\bm m_{l,r}\parallel \bm P_{l,r}$. Then  only the internal chirality (\ref{Eq:chi1}) is non-zero and it is given by
$\chi_1 = P \bm m\cdot (\bm n_\alpha\times \bm n_\theta) $ evaluated at left and right FM/SC interfaces. 
 The full expressions for non-zero anomalous $j_{in} \neq 0$ and ordinary current $j_{o} \neq 0$ are given in the Appendix \ref{Sec:Ap1}.
 The answer is especially simple for $T \to T_c$ and in the tunnel limit $\gamma k_\omega \gg 1$:
 %
 %
   \begin{align} 
   \label{Eq:current_simp}
 &  \frac{ ej_{o}}{\sigma_n} = \frac{2\Delta^2}{\pi T_c}(1-P^2) e^{-\frac{d}{\xi_N}}
   \frac{ \xi_h^2 \xi_N \theta^{\prime 2} 
 \sin^2 \! \alpha \cos[\theta^\prime d \cos \! \alpha]}{ \gamma^2}
 \\
 \label{Eq:current_in_simp}
 & \frac{j_{in}}{j_o}  = 2  P \frac{\xi_N}{\gamma} 
 \tan [\theta^\prime d \cos \! \alpha],~~~~~~~~~
   \end{align}
where $\xi_N = \sqrt{D/2\pi T_c}$. It is seen that in this case the anomalous current is an odd function of  $\theta^\prime \cos \alpha$. Therefore, it is determined by the internal chirality of the texture. In more general case it is determined by the quantity $(\chi_{1l}+\chi_{1r})$, which represents the combination of the internal chirality of the texture and the projection of the interface polarization on the local magnetization $(\bm P_l \bm m_l + \bm P_r \bm m_r)$.


Due to the presence of the anomalous current the state with $\varphi=0$ or $\varphi=\pi$ is no more the ground state of the system. The ground state phase difference is determined from the condition $j=0$ and takes the form $\tan \chi_0 = -j_a/j_o$.  In the limit $\theta' \ll k_\omega$, $\gamma k_\omega \gg 1$ and $T \to T_c$ the anomalous phase shift takes the simplest form
\begin{eqnarray}
\tan \varphi_0 = - \frac{2 P \xi_N}{\gamma}\tan(\theta' d\cos\alpha).
\label{chi_0}
\end{eqnarray}

It can be concluded that the anomalous phase shift is more pronounced for transparent junctions. It is only non-zero for non-coplanar magnetic texture $\alpha \neq \pi/2$ and is absent if the interface has no spin-filtering properties $P=0$ in accordance with the previous considerations of the Josephson current via the magnetic helix \cite{Volkov2006}.

 \subsubsection{Numerics}

 Here we consider the case when only the internal chirality is non-zero $\chi_1\neq 0$ while  $\chi_2 = \chi_3 = 0$. 
 Numerically we study the parameter regions, which are not covered by the above analytical treatment. In this section all energies are measured in units of $T_c$ and all lengths are measured in units of $\xi_N$. Therefore, $\theta'$ is measured in units of $\xi_N^{-1}$. Currents are measured in units of $(\Delta^2/16 \pi T_c)(\sigma_n/e \xi_N)$. 
 
 Typical values of $j_a/j_o$ are very small. Therefore the ground state phase difference is also very small. The exception are the parameter regions corresponding to vicinities of $0$-$\pi$ transitions in the Josephson junction, where the value of $j_o$ goes to zero and, consequently, the value of the anomalous ground state phase difference can have arbitrary values between $0$ and $\pi$.

 \begin{figure}[htb!]
 \centerline{\includegraphics[clip=true,width=3.4in]{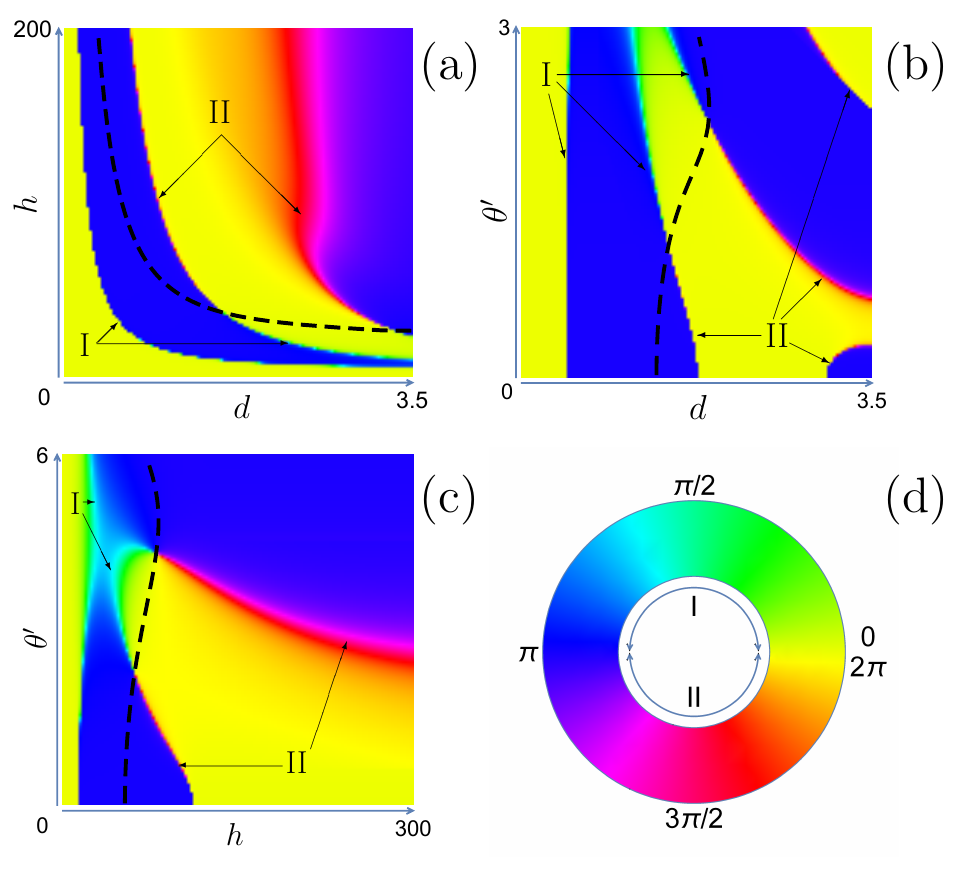}}
 \caption{\label{Fig:2D}  (Color online)
 (a) 2D map of the ground state phase in $(d,h)$-plane. 
  The black dashed line corresponds to $j_a=0$, $\theta'=1$,
 (b) 2D map of the ground state phase in $(d,\theta')$-plane, 
  $h=40$;
 (c) 2D map of the ground state phase in $(h,\theta')$-plane, $d=1$. 
  The other parameters are $P=0.5$, $\alpha=\pi/4$, $\gamma = 1$ for panels (a)-(c).
  (d) The color coding to panels (a-c).
       }
 \end{figure}   
 
In order to explore the ground state of the junction in detail we demonstrate the 2D color-coded plots of the ground state phase $\varphi_0$. Fig.~\ref{Fig:2D}(a) represents the phase diagram in $(d,h)$-plane. The color coding of the phase is explained in Fig.~\ref{Fig:2D}(d). One can see the regions of $0$-state (in yellow) and the regions of $\pi$-state (in blue). There are also regions of the intermediate ground state phase between them. 

The most striking feature, which can be observed from these plots is that there are two topologically inequivalent types of  $0$-$\pi$ transitions in the system. It is seen that the transitions between $0$ and $\pi$ states differ by the way of the unit circle bypass, see Fig.~\ref{Fig:2D}(d). The green-light blue regions (boundaries between yellow $0$-regions and blue $\pi$-regions) correspond to the transitions via $\pi/2$ intermediate phases and are called by type I transitions. The red boundaries indicate transitions via $3\pi/2$ intermediate phases and are called by type II transitions. The type of the transition is determined by the sign of $j_a$. All the phase diagrams in Fig.~\ref{Fig:2D} are plotted for $\chi_1 >0$. That is the type of the transition changes to the opposite everywhere in these phase diagrams for the opposite chirality. This statement is also valid when the dependence of the anomalous current on the phase difference $\varphi$ is more general and not restricted by $\cos \varphi$. 

The type of the transition can also be changed for a fixed chirality due to the alternating dependence of $j_a$ on the junction parameters: $h$, $|\theta^\prime|$ and $d$. It is demonstrated in Fig.~\ref{Fig:2D}(a), where 
the black dashed line corresponds to $j_a = 0$ and the intersections of this line with the transition lines $j_o = 0$ give the points where the transition type changes.

Figs.~\ref{Fig:2D}(b) and \ref{Fig:2D}(c) represent analogous color-coded 2D maps of the ground state phase difference
 $\varphi_0$ in planes $(d,\theta')$ and $(h,\theta')$, respectively. Together with Fig.~\ref{Fig:2D}(a) they provide the complete picture of the ground state phase distribution and the transition types in the system under consideration.

As it is described by Eq.~(\ref{energy_harmonics}), the anomalous contribution to the Josephson energy can be expressed via the anomalous current as $E_{ch}=(\hbar/2e)j_a \sin \varphi$. 
Fig.~\ref{Fig:energy_helix}c demonstrates the anomalous contribution $E_{ch}(\varphi = \pi/2)/E_{J0}$, where $E_{J0}=(\hbar/2e)\sqrt{j_o^2 + j_a^2}$, as a function of $|\theta^\prime|$ for different $d$. Black and red lines correspond to different chiralities ${\rm sgn}\chi_1 = \pm 1$. Therefore, for the particular example of the magnetic helix this figure clearly illustrates our statement that the chiral contribution to the Josephson energy removes the degeneracy between opposite chiralities.

 \begin{figure*}[htb!]
 \centerline{$
 \begin{array}{c}
 \includegraphics[width=1.0\linewidth]{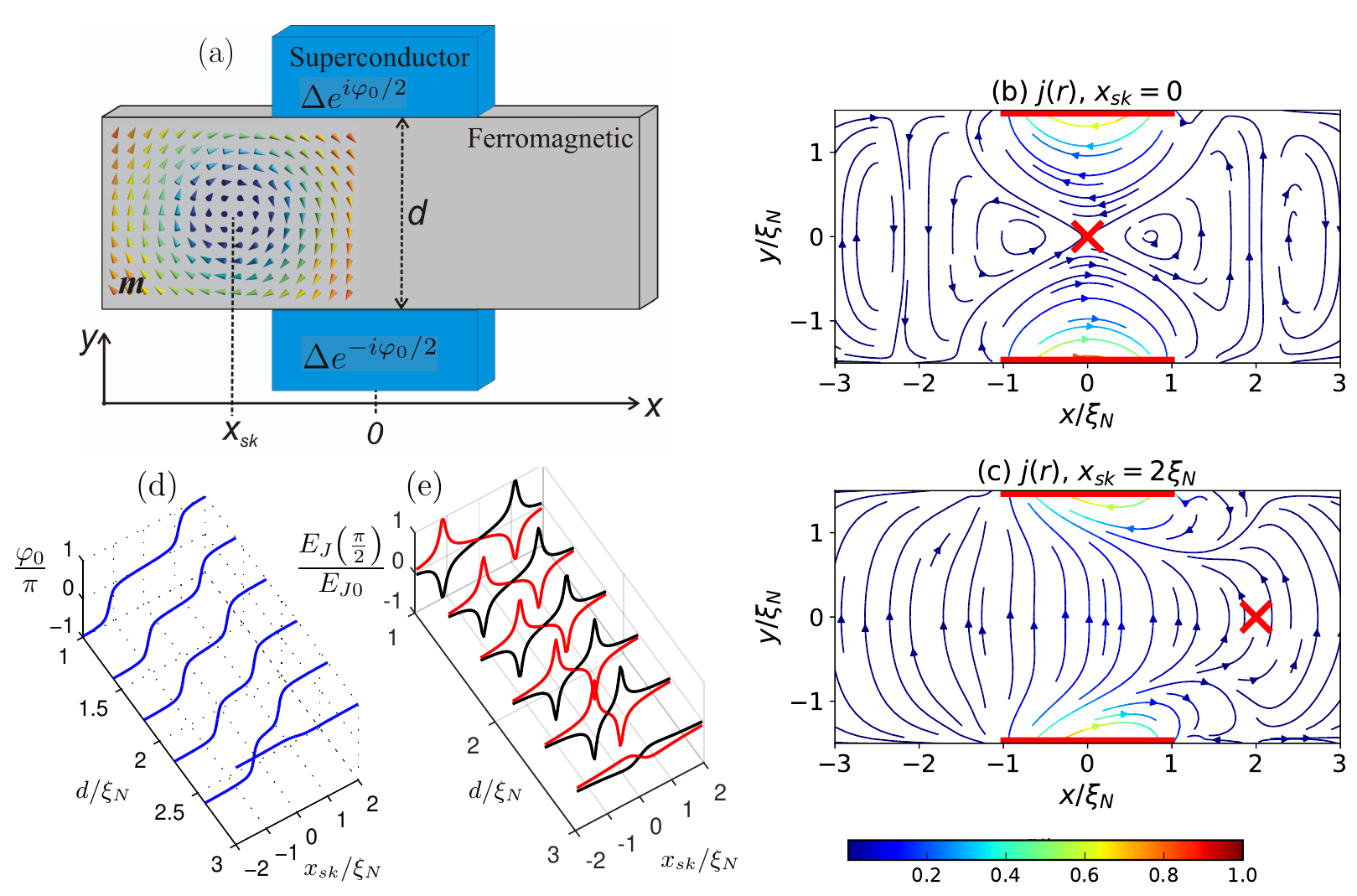}
  \end{array}$}
 \caption{\label{Fig:SFS}
 (a) SC/FM/SC junction through the ferromagnetic tape  skyrmion magnetic texture.
 (b,c) Normalized current densities ${\bm j}$ current induced in the ferromagnetic layer for the SFS shown in the panel (a) with $\varphi=0$.
 The skyrmion center marked with red cross is located at (b) $x_{sk}=0$, $y_{sk}=0$ and (c) $x_{sk}= 2\xi_N$ and $y_{sk}=0$. 
The exchange field is $h=5T_c$ and temperature $T=0.5T_c$. 
 The SC/FM interfaces  at $y=\mp 1.5 \xi_N$ are marked by the thick red lines. 
  (d) The ground state phase difference as the functions of the skyrmion positions with $Q=1$ for  different values of the junction width $d$.  (e) Anomalous Josephson energy $E_J(\varphi=\pi/2)$. 
Red and black curves correspond to the different skyrmion topological charges $Q= 1$ and $Q=-1$, respectively. 
}
 \end{figure*}

 \subsection{Magnetic skyrmion}

As it was already discussed in the introduction, the situation with the anomalous current and intermediate ground state phase is not rare in Josephson junctions based on S/F hybrids. The only essential condition is the noncoplanarity of the magnetization in the system. If this condition is fulfilled, when the ground state phase difference should, in general, have intermediate values if the exchange field is treated beyond the framework of the quasiclassical theory. Below we consider another example, namely the Josephson junction through a magnetic skyrmion. We demonstrate the possibility of skyrmion manipulation and detection using the effect of anomalous Josephson  energy term resulting from the triple spin interactions.  
    
 Magnetic skyrmions are the topological spin textures that can be 
 spontaneously formed in magnetic systems due to the various mechanisms \cite{Nagaosa-NatureNano2013}. 
 Recent interest to these objects has been stimulated by the 
 discovery of skyrmions stabilized by chiral interactions in ferromagnets  
 with broken spatial inversion symmetry 
 \cite{Roszler2006,Nagaosa-NatureNano2013,Muehlbauer2009,Yu2010}.
 Owning to their small size and high mobility 
 using skyrmions instead of domain walls\cite{Fert2013} has been suggested as a possible way to significantly 
 improve the performance of magnetic racetrack memory architectures\cite{Parkin2008}.
 Being much less sensitive to the defect pinning 
 skyrmions can be manipulated via the spin-transfer torque under the ultra-small current 
 densities\cite{Jonietz2010,Yu2012,Schulz2012}.

 The example of SF system featuring ground-state currents
 consists of the magnetic skyrmion surrounded by superconducting material as 
 is shown schematically in Fig.(\ref{Fig:SFS}a). Here arrows depict the direction of magnetization
 ${\bm m}$ at a given point described by the general expression (\ref{Eq:Skyrmion}) 
 with vorticity $\kappa_v=1$, helicity $\nu=1$ and the radial distribution $\cos\Theta = (r^2-\xi_N^2/4)/(r^2+\xi_N^2/4) $. 

 In Fig.(\ref{Fig:SFS}a) a skyrmion with positive topological charge $Q=1$ is shown.
We study the anomalous Josephson effect in this setup and 
  show that {the anomalous Josephson energy given by the third term in (\ref{energy_harmonics})} removes the 
   degeneracy of $Q=\pm 1$ states, provided that  the Josephson phase difference is kept fixed. 
   That is, depending  on the sign of $j_a$ amplitude either the skyrmions with $Q=+1$ or $Q=-1$ with the fixed vorticity $\kappa_v$  
   become energetically cheaper. 
   This feature makes the proposed energy contribution
   qualitatively different from the DM interaction which although selects the vorticity and helicity values but 
   does not fix the polarity and therefore {allows for} the overall sign change of $Q$ under the global magnetization inversion.

  From the general symmetry arguments one can see that the direction of spontaneous current 
  shown in Fig.\ref{Fig:SFS}(b,c) is determined by the topological number $Q$ but is not sensitive to the value of helicity $\nu$.
  To demonstrate that one can consider the $\pi$-rotation of the whole SC/FM system with skyrmion texture shown in 
  Fig.(\ref{Fig:SFS}a), around the $x$-axis. This transformation flips the direction of circulating current ${\bm j}$ 
  together with the sign of $Q$. The invariance with respect to helicity $\nu$ change can be understood by looking at the system reflected in 
  $xy$ plane. The mirror images of charge currents circulating in $xy$ plane keep the same direction. 
  At the same time this transformation flips the signs of $m_{x,y}$ components  resulting in the sign change of $\nu$ while the component $m_z$ together with the topological charge $Q$ remains intact. 
     
 To illustrate the above general arguments we describe
 weak proximity effect in the ferromagnetic layer by the linearized Usadel theory
 (\ref{Eq:Usadel},\ref{bc},\ref{Eq:Current1}). 
 The spin-filtering effect is determined by the barrier polarization ${\bm P}$ with the amplitude 
 $0<P<1$. Here in order to keep the topology of magnetic configuration we assume that 
 barrier polarization is parallel to the local value of the exchange field at the SC/FM interface ${\bm P}\parallel {\bm m}$.
 Hence the sign of $P_z$ projection is determined by the topological charge: $P_z<0$ for $Q=1$ and $P_z>0$ for $Q=-1$.
  
  In order to find spontaneous currents in this system we solved numerically the linear 
  boundary value problem (\ref{Eq:Usadel},\ref{bc},\ref{Eq:Current1}) with  
  magnetization distribution (\ref{Eq:Skyrmion}). We consider different widths of the ferromagnetic layer $d$ and the shift 
  of magnetic skyrmion $x_{sk}$ with respect to the superconducting electrodes.  
  We have solved the system of linear partial differential equations at the finite element framework\cite{Hecht2012}.
  
  The examples of supercurrent distributions 
  at two different skyrmion positions marked by the red cross are shown in Figs.(\ref{Fig:SFS}b,c)
  Current density here is normalized by its maximal value reached at the SC/FM interfaces. One can see that the  skyrmion shifted from the center to $x_{sk}\neq 0$ generates net current between superconducting electrodes. For the skyrmion at the geometrical symmetry point $x_{sk}=y_{sk}=0$ the current density is finite but the net current is absent.  
 
 We study the net current as function of parameters $d,x_{sk}$,
 where 
 shifting the skyrmion along the line $y_{sk}=0$.  
  The CPR obtained within the linearized theory have the exact form (\ref{current_harmonics}) without admixture of 
  higher harmonics. The anomalous current generates the spontaneous phase shift $\varphi_0$.      
 The behavior of the ground state phase shift $\varphi_0(d,x_{sk})$ generated by the skyrmion with topological charge $Q=1$ is shown in Fig.\ref{Fig:SFS}d as a function
  of the width of ferromagnetic layer $d$ and the skyrmion shift $x_{sk}$ along the junction.  
  First of all one can see that the non-trivial state with $\varphi_0\neq \pi n$ is possible and in fact is a rather 
  generic one.  It exists elsewhere in the parameter space except for the symmetry point $x_{sk}=y_{sk}=0$. 
  In this case the system has the magnetization-inversion symmetry determined by the real space $\pi$-rotation around 
  the $x$
  axis. As we discussed above in this case $\varphi_0$ state is absent.  
   
 Of particular interest is the possibility to realize a tunable $\varphi_0$-junction where 
 we can set an arbitrary equilibrium phase difference by shifting the skyrmion within the spacer between superconducting leads. 
 In Fig.\ref{Fig:SFS}(d) one can see {that  the system} demonstrates the possibility to obtain 
 an arbitrary value of the ground state phase near the $0-\pi$ crossover as a function of $d$ and $x_{sk}$. 
 Comparing Figs.\ref{Fig:2D} and \ref{Fig:SFS}d  one can see that 
 this behavior is similar to that obtained in the setup with magnetic helix considered above. 
 
 The  Josephson energy at the fixed phase difference $\varphi=\pi/2$ is shown in Fig.\ref{Fig:SFS} for $Q=\pm 1$.
 One can see that this energy removes the degeneracy of $Q=\pm 1$ states. 
For the fixed phase difference and in the absence of the pinning forces  the minimum Josephson energy determines the equilibrium position of the skyrmion with respect to the superconducting electrodes.
Thus, the equilibrium position of skyrmion is determined by the Josephson energy minimum and can be controlled by the tuning the phase difference. With  increasing  phase difference from $0$ to $\pi/2$ the skyrmion shifts from the center to the energy minimum coordinate shown in  Fig.\ref{Fig:SFS}d.

 \section{Concluding remarks}
 \label{Sec:discussion}

 Besides fundamental importance the suggested 
 spin interaction mechanism can have several practical applications.  
 Recently the current-driven  skyrmion dynamics has attracted large interest as the possible 
route to  low-power manipulation of magnetic textures \citep{Nagaosa2013,Fert2013}. First, one can use the anomalous Josephson effect for the fast detection of  skyrmions moving along the ferromagnetic tape in a skyrmion racetrack memory design,\cite{Parkin2008,Fert2013}.  
Such superconducting skyrmion detector can be realized using the system geometry shown schematically in Fig. (\ref{Fig:SFS})a. 
At the fixed Josephson current, e.g. $j=0$ the phase difference across the junction  depends on the skyrmion position through $\varphi = \varphi_0(x_{sk})$ so that moving skyrmion should generate the voltage pulse $U_J(t)$ between superconducting electrodes. In the limit of large junction  resistances
 one can neglect the normal current contribution and estimate this voltage as  $U_J = - \hbar \partial_t\varphi_0 /2e $ where $\partial_t\varphi_0 = \partial_t x_{sk}\partial_{x_{sk}} \varphi_0 $.  Using this effect it is possible to detect the individual skyrmions passing through the Josephson junction while moving along the ferromagnetic layer. 

The inverse effect can be used for moving skyrmions with the help of dissipationless superconducting current.
As we discussed above for the setup in Fig.(\ref{Fig:SFS})a, the equilibrium skyrmion position relative to the superconducting electrodes depends on the Josephson phase difference. 
This effect is fact determined by the finite width of superconducting electrodes, which e.g. is equal to $2\xi_N$
for the geometry used for producing the results in Figs.(\ref{Fig:SFS})b-e. If  SC electrodes are very wide the system can be considered as translational invariant along the $x$-axis. In this case, provided we can neglect the pinning force there is no equilibrium  position of skyrmion and it moves continuously along $x$ axis with the drifting velocity determined by the balance of the effective spin-torque term and/or Gilbert damping.  The force acting on the skyrmion from the supercurrent results from the adiabatic spin torque mechanism. Its analytical expression can be obtained in case of the strong ferromagnet using the formalism of extended quasiclassical theory \cite{PhysRevB.98.014521}. 
The anomalous chirality-selective energy contribution  results in  the force having opposite  directions for skyrmions with $Q=\pm 1$.

The considered examples demonstrate that triple spin interaction energetically prefers one of magnetic textures with opposite chiralities, otherwise degenerated. In general, such interaction arises in noncoplanar magnetic textures. Therefore, it is completely different from the situation in thin magnetic films \cite{Heide2008,Thiaville2012,Emori2013}, where left-handed or right-handed plane Neel domain walls are preferred by the combination of magnetostatic and DM energies. As opposed to the considered here noncoplanar textures, for the plane textures it is not possible to ascribe a definite chirality defined as a mixed product $\bm m_1\cdot (\bm m_2 \times \bm m_3)$ to a particular texture. It is only possible to distinguish between left-handed and right-handed textures. Due to the absence of this chiral invariant the left-handed and right-handed plane textures are still degenerated with respect to the global magnetization inversion.

In the present paper we have demonstrated that the widely known anomalous Josephson effect via a noncoplanar magnetic trilayer structure is a particular manifestation of a general triple spin interaction mechanism, which works for any noncoplanar magnetic system beyond the Josephson field. We would like also to note that in Refs.~\onlinecite{Liu2010,Kulagina2014} it was concluded that for the trilayer magnetic interlayer the noncoplanarity by itself is not enough to obtain the anomalous Josephson effect. It was claimed that the internal scattering barriers play an instrumental role in creating this effect. We believe that it is a consequence of a particular choice of a model system to investigate. By considering the simple example of three magnetic moments we demonstrate that the only necessary condition for this interaction is noncoplanarity of the magnetic texture.

To summarize we have introduced spin interaction which is fundamentally different from the previous mechanisms. It is generated due to the indirect exchange mediated by the moving superconducting condensate, modulating the spin response of  the conductivity electrons either in the superconductor hosting magnetic impurities or the hybrid superconductor/ferromagnet structures with the proximity-induced superconducting correlations. The generic example of three magnetic impurities demonstrates the origin and magnitude of this effect. The realistic Josephson devices with magnetic helix and skyrmion 
provide the motivation for the future experimental and practical applications. Possible advances in spintronics effect which  direction is  based on the low-dissipative manipulation and detection of skyrmions in the Josephson racetrack geometry suggested in Fig.\ref{Fig:SFS}.  

 \section{Acknowledgements}
 We thank Sebastian Bergeret, Ilya Tokatly, Tero Heikkila and Alexander Mel'nikov for stimulating discussion. We used  Matplotlib\cite{Hecht2012} package to produce plots.
The work of M.A.S. was supported by the Academy of Finland.  
 We also acknowledge the financial support by the Russian-Greek project No. 2017-14-588-0007-011 "Experimental and theoretical studies of physical properties of low-dimensional nanoelectronic systems" (D.S.R, I.V.B. and A.M.B.) and the RFBR project 18-52-45011 (I.V.B. and A.M.B.).
 
 \appendix
 \section{Derivation of the anomalous current-phase relations through magnetic helix} \label{Sec:Ap1}
  
  To find the amplitudes $f_\alpha$ and $f_\theta$ we project the Usadel equation (\ref{Eq:Usadel}) to the orthogonal 
vectors $\bm n_\theta$ and $\bm n_\alpha $ to obtain
 {
 \begin{align} \nonumber
 &  (\nabla^2 - \theta^{\prime 2} -k_\omega^2) f_\theta + 2  (\theta^\prime \cos \alpha)^2 \nabla f_\alpha =0
 \\
 \nonumber
 &  (\nabla^2 - \cos^2 \alpha  \theta^{\prime 2}-k_\omega^2) f_\alpha - 
 2  \nabla f_\theta =0 ,
 \end{align}
 where we denote $k_\omega =  \sqrt{2|\omega|/D}$.
 Thus we get the long-range modes $f_{\alpha,\theta}\propto e^{\zeta x}$, 
 where $\zeta$ is given by 
 \begin{equation}
 (\zeta^2-k_\omega^2 - \theta^{\prime 2})(\zeta^2-k_\omega^2 - \theta^{\prime 2} \cos^2 \alpha ) + (2\cos \alpha \theta^{\prime}\zeta)^2 =0,
 \end{equation}
 which in the limit of $\theta^{\prime}\gg k_\omega$ is reduced to the one obtained in Ref.~\onlinecite{Volkov2006}
 \begin{equation}
 \zeta^4 - \zeta^2 \theta^{\prime 2} (1-3\cos^2 \alpha) + \cos^2 \alpha \theta^{\prime 4} =0.
 \end{equation}
 }

 
{\bf Coupling short and long range modes at the interface.}
First, we determine the solution for short-range modes in the form of (\ref{Eq:SRSC0})-(\ref{Eq:SRTC0}). 
 The coefficients can be found from the boundary conditions projected on the $\bm m$ direction
 \begin{align}
 & \left(\partial_n + \frac{G_s}{\gamma}\right) f_{sr} + \theta^\prime \sin^2 \alpha f_\theta = 
 \frac{iG_s}{\gamma}\Bigl[\chi_2 f_\theta + \chi_3 f_\alpha \Bigr]
 \\
 & \left(\partial_n + \frac{G_s}{\gamma}\right) f_0 =  \frac{\sqrt{1-P^2}}{\gamma} F_{S}, 
 \end{align}
 where the chiralities $\chi_{2,3}$ are given by the Eqs.(\ref{Eq:chi2},\ref{Eq:chi3}).
 
 As it was already mentioned $\theta^\prime \ll k_h$. Further we also assume $G_s/\gamma \ll k_h$. In this case 
 the coefficients are given by
 \begin{align} \label{Eq:X1sr}
 & X_1= \sqrt{1-P^2}F_{Sl}/(2\gamma\lambda) 
 \\ \label{Eq:X2sr}
 & X_2= \sqrt{1-P^2}F_{Sl}/(2\gamma\lambda^*) 
 \end{align}
 Here we consider the vicinity of $x=0$ SC/FM interface and the opposite boundary at  $x=d$ can be described by changing  $x\to d-x$ and $F_{Sl} \to F_{Sr}$. 

 The boundary conditions that provide coupling between the long-range and short-range modes 
 are obtained directly from (\ref{bc}) and read in components 
   \begin{align}
  \label{Eq:long_theta}
  & \sin^2 \! \alpha(\partial_n + \frac{G_s}{\gamma})f_\theta  = 
  (\pm \theta^\prime \sin^2 \alpha - \frac{iG_s}{\gamma}\chi_2) f_{sr} \mp A f_\alpha 
  \\ \label{Eq:long_alpha}
  & \frac{1}{4}\theta^{\prime 2}\sin^2 2 \alpha(\partial_n + \frac{G_s}{\gamma}) f_\alpha = 
  \pm A f_\theta - \frac{iG_s}{\gamma}\chi_3 f_{sr}
  \\ \label{Eq:A}
  & A= -\frac{1}{4}\theta^{\prime 2}(\sin 2 \alpha)^2 \pm i\gamma^{-1} \chi_1 G_S
  \end{align}
 where the upper and lower signs describe $x=0$ and $x=d$ interfaces, respectively. Upon writing Eqs.~(\ref{Eq:long_theta})-(\ref{Eq:long_alpha}) we take into account that the short-range modes also have components $f_{\theta,sr}$ and $f_{\alpha,sr}$ along $\bm n_\theta$ and $\bm n_\alpha$ directions. They are small by a factor of $\theta'/k_h$ with respect to the components written in Eqs.~(\ref{Eq:SRSC0}) and (\ref{Eq:SRTC0}), but their spatial derivatives should be accounted for in Eqs.~(\ref{Eq:long_theta})-(\ref{Eq:long_alpha}) and can be obtained by integrating the Usadel equation over the spatial region $0<x<x^*$ near the interface, where $\xi_h \ll x^* \ll \xi_N$. This procedure gives us $\partial_x f_{\theta,sr}=2 \theta^\prime f_{sr}$. The short-range triplet amplitude is determined from (\ref{Eq:SRTC0}) as $f_{sr}=i\gamma^{-1}\sqrt{1-P^2} F_{S} {\rm Im} \lambda^{-1}$, where $F_{S} = e^{\mp i\varphi/2} \Delta/\sqrt{\Delta^2 +\omega^2}$ is the anomalous function in the left (right) electrode.   
 
 To simplify the analytical calculations we consider the 
  case when the distance between SC electrodes is larger than the decay lengths of long-range solutions  $\zeta_k d>1$. 
  Then the long-range solutions generated at $x=0$ 
  can be found in the form
 \begin{align}
 \label{Eq:f_theta}
 & f_\theta =  f_{sr}( C_1 e^{-\zeta_1 x} + C_2 e^{-\zeta_2 x} )
 \\\label{Eq:f_alpha}
 & f_\alpha = f_{sr} (a_1 C_1e^{-\zeta_1 x} + a_2 C_2e^{-\zeta_2 x} )
 \end{align}
 where 
 $a_k = (\zeta_k^2 - \theta^{\prime 2} - k_\omega^2)/(2\theta^{\prime 2} (\cos \alpha)^2 \zeta_k)$. The long-range solutions generated at $x=d$ interface are obtained from Eqs.~(\ref{Eq:f_theta})-(\ref{Eq:f_alpha}) by $a_k \to - a_k$ and $x \to d-x$.
 
 Coefficients $C_{1,2}$ are to be found from the boundary conditions Eqs.~(\ref{Eq:long_theta})-(\ref{Eq:long_alpha}) 
 and take the form: 
  \begin{widetext}
  \begin{align}
 & C_1 = \frac{1}{Z}\Bigl\{ \left[ -A + \frac{1}{4}\theta^{\prime 2}
 \sin^2 \! 2\alpha \left(\zeta_2 + \frac{G_s}{\gamma} \right)a_2 \right] 
 \left[ \pm \theta^\prime \sin^2 \! \alpha - i \frac{G_s}{\gamma} \chi_2 \right]\pm 
  i \frac{G_s}{\gamma}\chi_3 \left[ \sin^2\!\alpha \left(\zeta_2 + \frac{G_s}{\gamma}\right) + a_2 A \right] \Bigr\}
   \nonumber
  \\
 & Z=(a_2 - a_1)\bigl[ A^2 + \theta^{\prime 2}\sin^4 \! \alpha \cos^2 \! \alpha \left(\zeta_1 + \frac{G_s}{\gamma}\right) 
 \left(\zeta_2 + \frac{G_s}{\gamma}\right) \bigr] + 
  A \sin^2 \! \alpha (\zeta_2 - \zeta_1)\bigl[  a_1 a_2 \theta^{\prime 2}\cos^2 \! \alpha + 1 \bigr]~~~~~~~~
  \label{Eq:C1}
 \end{align}
 \end{widetext}
 with similar expression for $C_2$ obtained by the symmetric interchange of $a_1 \leftrightarrow a_2$, $\zeta_1 \leftrightarrow \zeta_2$ and $A  $ given by (\ref{Eq:A}).

 {\bf Current in the long junction $d>\xi_N$.} 

Now our goal is to find the current-phase relation.
In case $d>\xi_N$ in the middle of the interlayer the current is transmitted by long-range modes, and we can rewrite Eq.~(\ref{Eq:Current1}) as
\begin{align}
\frac{ej}{2\pi T \sigma_n} = -\sin^2 \! \alpha \sum \limits_{\omega > 0} {\rm Im} \Bigl[ f_\theta^* \partial_x f_\theta + \nonumber
\\
\theta^{\prime 2}\cos^2\! \alpha (f_\alpha^* \partial_x f_\alpha + 2 f_\theta^* f_\alpha ) \Bigr] 
\label{Eq:current_long}
\end{align}

Further we analyze the limit of 
"slow" magnetic helix $\theta^\prime \ll k_\omega$. It means that the helix magnetization rotates slowly on the scale $\xi_N$. In this case $\zeta_1 = \zeta_2^* = k_\omega + i \theta^\prime \cos \! \alpha$ and Eq.~(\ref{Eq:current_long}) can be written as
\begin{align}
\frac{ej}{2\pi T \sigma_n} = 2 \sin^2 \! \alpha \sum \limits_{\omega > 0} {\rm Im} \Bigl[ e^{-\zeta_1 d} (C_{1l}C_{2r}^*e^{-i\varphi} -C_{1r}C_{2l}^* e^{i\varphi}) \times \nonumber 
\\
\bigl\{ \zeta_1 - \theta^{\prime 2}\cos^2\! \alpha (a_1^2 \zeta_1 + 2 a_1) \bigr\}\Bigr]|f_{sr}|^2,~~~~~~~~~~~~ 
\label{Eq:current_limit}
\end{align}
where $C_{1(2)l}$ and $C_{1(2)r}$ are determined by Eq.~(\ref{Eq:C1}) at $x=0$ and $x=d$ interfaces, respectively.

The resulting CPR takes the form of Eq.~(\ref{current_harmonics}) with anomalous $j_a$ current contribution
given by the superposition of three parts $j_a = j_{ex} + j_{in} + j_{mix}$. Together with the ordinary Josephson current amplitude $j_o$ they are given by 
 \begin{widetext}
 \begin{align}
  & \frac{ej_{in}}{4\pi T \sigma_n} = 
 \bigl(\chi_{1l}+\chi_{1r})
  \frac{ \theta^{\prime}\sin (\theta^\prime d \cos \! \alpha )}{ \cos \! \alpha }
 \sum \limits_{\omega > 0}  \left(\frac{G_s}{\gamma}\right)  \frac{ k_\omega \tilde k_\omega }{K} 
 e^{-k_\omega d}|f_{sr}|^2  
 ~~~~~~~~~~~~~
 \label{Eq:current_in}
 \\
 & \frac{ej_{ex}}{4\pi T \sigma_n} = 
  (\bm P_l \bm m_{l}+\bm P_r \bm m_{r})
 ( \chi_{2l}\chi_{3r}-\chi_{3l}\chi_{2r})
 \frac{\cos (\theta^\prime d \cos \! \alpha )}{\theta^\prime \cos \! \alpha \sin^2 \! \alpha}
 \sum \limits_{\omega > 0} \left( \frac{G_s}{\gamma } \right)^3
 \frac{ k_\omega \tilde k_\omega }{K} 
   e^{-k_\omega d}|f_{sr}|^2 
  \label{Eq:current_ex}
 \\
 & \frac{ej_{mix}}{4\pi T \sigma_n} = - 
       \bigl(\chi_{1l}+\chi_{1r})  
 \left[\chi_{2l}\chi_{2r}+\frac{\chi_{3l}\chi_{3r}}{\theta^{\prime 2} \cos^2 \! \alpha }\right]  
 \frac{\sin (\theta^\prime d \cos \! \alpha )}{\theta^\prime \cos \! \alpha \sin^4 \! \alpha}
 \sum \limits_{\omega > 0}   \left(\frac{G_s}{\gamma}\right)^3  \frac{ k_\omega \tilde k_\omega }{K} 
   e^{-k_\omega d}|f_{sr}|^2 
 \label{Eq:current_mix}
 \\
 & \frac{ej_o}{4\pi T \sigma_n} = \sum \limits_{\omega > 0} \frac{k_\omega}{K}    e^{-k_\omega d}|f_{sr}|^2 
\frac{ \tilde k_\omega^2 + \bigl( \frac{G_s}{\gamma} \bigr)^2(\bm P_l \bm m_l)
(\bm P_r \bm m_r)}{\sin^2 \! \alpha} 
\\
& \Bigl[ \cos[\theta^\prime d \cos \! \alpha]\bigl\{\sin^4 \! \alpha \theta^{\prime 2} - \left( \frac{G_s}{\gamma} \right)^2 (\chi_{2l} \chi_{2r} + \frac{\chi_{3l} \chi_{3r}}{\theta^{\prime 2} \cos^2 \! \alpha}) \bigr\} + 
\left(\frac{G_s}{\gamma}\right)^2\frac{\sin[\theta^\prime d \cos \! \alpha]}{\theta^\prime \cos \alpha} (\chi_{2r} \chi_{3l}-\chi_{2l} \chi_{3r})\Bigr].~~~~~~~~~~~~
\label{Eq:current_normal}
 \end{align}

 \end{widetext}
  where $\bm m_{l} = \bm m (0)$ and $\bm m_{r} = \bm m (d)$ are the magnetizations at the FM/SC interfaces, 
$ K= \left [( \tilde k_\omega^2 -  (G_s\bm P_l \bm m_{l}/\gamma)^2 \right] 
 \left[( \tilde k_\omega^2 -  (G_s \bm P_r \bm m_{r}/\gamma)^2 \right] $
  and $\tilde{k}_\omega = k_\omega + G_s/\gamma$.

  
In case when $\theta^\prime \neq 0$ 
but 
$\bm m_{l,r}\parallel \bm P_{l,r}$ only the internal chirality (\ref{Eq:chi1}) is non-zero and it is given by
$\chi_1 = P \bm m (\bm n_\alpha\times \bm n_\theta) $. 
Then the only nonzero contribution to the anomalous current is $j_{in}$, which takes the form
\begin{align}
\frac{ej_{in}}{2\pi T \sigma_n} = \frac{8 \Delta^2 P (1-P^2)\sin[\theta^\prime d \cos \! \alpha] \theta^{\prime 2}\sin^2 \! \alpha}{k_h^2} \times \nonumber 
\\
\sum \limits_{\omega > 0} \Bigl( \frac{G_s}{\gamma} \Bigr)^3 \frac{k_\omega \tilde k_\omega ^{-k_\omega d}}
{\omega^2 \left[ \tilde k_\omega^2 - \Bigl( PG_s/\gamma \Bigr)^2   \right]^2}
\label{Eq:current_in_final}
\end{align}
The answer is especially simple for $T \to T_c$ and in the tunnel limit $\gamma k_\omega \gg 1$ we get the Eq.(\ref{Eq:current_in_simp}).
 In the same limit the ordinary Josephson current amplitude (\ref{Eq:current_normal}) reduces to the simpler expression (\ref{Eq:current_simp}). 

%

 \bibliography{refs}

\begin{thebibliography}{104}%
\makeatletter
\providecommand \@ifxundefined [1]{%
 \@ifx{#1\undefined}
}%
\providecommand \@ifnum [1]{%
 \ifnum #1\expandafter \@firstoftwo
 \else \expandafter \@secondoftwo
 \fi
}%
\providecommand \@ifx [1]{%
 \ifx #1\expandafter \@firstoftwo
 \else \expandafter \@secondoftwo
 \fi
}%
\providecommand \natexlab [1]{#1}%
\providecommand \enquote  [1]{``#1''}%
\providecommand \bibnamefont  [1]{#1}%
\providecommand \bibfnamefont [1]{#1}%
\providecommand \citenamefont [1]{#1}%
\providecommand \href@noop [0]{\@secondoftwo}%
\providecommand \href [0]{\begingroup \@sanitize@url \@href}%
\providecommand \@href[1]{\@@startlink{#1}\@@href}%
\providecommand \@@href[1]{\endgroup#1\@@endlink}%
\providecommand \@sanitize@url [0]{\catcode `\\12\catcode `\$12\catcode
  `\&12\catcode `\#12\catcode `\^12\catcode `\_12\catcode `\%12\relax}%
\providecommand \@@startlink[1]{}%
\providecommand \@@endlink[0]{}%
\providecommand \url  [0]{\begingroup\@sanitize@url \@url }%
\providecommand \@url [1]{\endgroup\@href {#1}{\urlprefix }}%
\providecommand \urlprefix  [0]{URL }%
\providecommand \Eprint [0]{\href }%
\providecommand \doibase [0]{http://dx.doi.org/}%
\providecommand \selectlanguage [0]{\@gobble}%
\providecommand \bibinfo  [0]{\@secondoftwo}%
\providecommand \bibfield  [0]{\@secondoftwo}%
\providecommand \translation [1]{[#1]}%
\providecommand \BibitemOpen [0]{}%
\providecommand \bibitemStop [0]{}%
\providecommand \bibitemNoStop [0]{.\EOS\space}%
\providecommand \EOS [0]{\spacefactor3000\relax}%
\providecommand \BibitemShut  [1]{\csname bibitem#1\endcsname}%
\let\auto@bib@innerbib\@empty
\bibitem [{\citenamefont {Ruderman}\ and\ \citenamefont
  {Kittel}(1954)}]{PhysRev.96.99}%
  \BibitemOpen
  \bibfield  {author} {\bibinfo {author} {\bibfnamefont {M.~A.}\ \bibnamefont
  {Ruderman}}\ and\ \bibinfo {author} {\bibfnamefont {C.}~\bibnamefont
  {Kittel}},\ }\href {\doibase 10.1103/PhysRev.96.99} {\bibfield  {journal}
  {\bibinfo  {journal} {Phys. Rev.}\ }\textbf {\bibinfo {volume} {96}},\
  \bibinfo {pages} {99} (\bibinfo {year} {1954})}\BibitemShut {NoStop}%
\bibitem [{\citenamefont {Kasuya}(1956)}]{doi:10.1143/PTP.16.45}%
  \BibitemOpen
  \bibfield  {author} {\bibinfo {author} {\bibfnamefont {T.}~\bibnamefont
  {Kasuya}},\ }\href {\doibase 10.1143/PTP.16.45} {\bibfield  {journal}
  {\bibinfo  {journal} {Progress of Theoretical Physics}\ }\textbf {\bibinfo
  {volume} {16}},\ \bibinfo {pages} {45} (\bibinfo {year} {1956})}\BibitemShut
  {NoStop}%
\bibitem [{\citenamefont {Yosida}(1957)}]{PhysRev.106.893}%
  \BibitemOpen
  \bibfield  {author} {\bibinfo {author} {\bibfnamefont {K.}~\bibnamefont
  {Yosida}},\ }\href {\doibase 10.1103/PhysRev.106.893} {\bibfield  {journal}
  {\bibinfo  {journal} {Phys. Rev.}\ }\textbf {\bibinfo {volume} {106}},\
  \bibinfo {pages} {893} (\bibinfo {year} {1957})}\BibitemShut {NoStop}%
\bibitem [{\citenamefont {Anderson}(1959)}]{PhysRev.115.2}%
  \BibitemOpen
  \bibfield  {author} {\bibinfo {author} {\bibfnamefont {P.~W.}\ \bibnamefont
  {Anderson}},\ }\href {\doibase 10.1103/PhysRev.115.2} {\bibfield  {journal}
  {\bibinfo  {journal} {Phys. Rev.}\ }\textbf {\bibinfo {volume} {115}},\
  \bibinfo {pages} {2} (\bibinfo {year} {1959})}\BibitemShut {NoStop}%
\bibitem [{\citenamefont {Moriya}(1960)}]{PhysRev.120.91}%
  \BibitemOpen
  \bibfield  {author} {\bibinfo {author} {\bibfnamefont {T.}~\bibnamefont
  {Moriya}},\ }\href {\doibase 10.1103/PhysRev.120.91} {\bibfield  {journal}
  {\bibinfo  {journal} {Phys. Rev.}\ }\textbf {\bibinfo {volume} {120}},\
  \bibinfo {pages} {91} (\bibinfo {year} {1960})}\BibitemShut {NoStop}%
\bibitem [{\citenamefont {Dzyaloshinsky}(1958)}]{DZYALOSHINSKY1958241}%
  \BibitemOpen
  \bibfield  {author} {\bibinfo {author} {\bibfnamefont {I.}~\bibnamefont
  {Dzyaloshinsky}},\ }\href {\doibase
  https://doi.org/10.1016/0022-3697(58)90076-3} {\bibfield  {journal} {\bibinfo
   {journal} {Journal of Physics and Chemistry of Solids}\ }\textbf {\bibinfo
  {volume} {4}},\ \bibinfo {pages} {241 } (\bibinfo {year} {1958})}\BibitemShut
  {NoStop}%
\bibitem [{\citenamefont {Crépieux}\ and\ \citenamefont
  {Lacroix}(1998)}]{CREPIEUX1998341}%
  \BibitemOpen
  \bibfield  {author} {\bibinfo {author} {\bibfnamefont {A.}~\bibnamefont
  {Crépieux}}\ and\ \bibinfo {author} {\bibfnamefont {C.}~\bibnamefont
  {Lacroix}},\ }\href {\doibase https://doi.org/10.1016/S0304-8853(97)01044-5}
  {\bibfield  {journal} {\bibinfo  {journal} {Journal of Magnetism and Magnetic
  Materials}\ }\textbf {\bibinfo {volume} {182}},\ \bibinfo {pages} {341 }
  (\bibinfo {year} {1998})}\BibitemShut {NoStop}%
\bibitem [{\citenamefont {Nagaosa}\ and\ \citenamefont
  {Tokura}(2013{\natexlab{a}})}]{Nagaosa2013}%
  \BibitemOpen
  \bibfield  {author} {\bibinfo {author} {\bibfnamefont {N.}~\bibnamefont
  {Nagaosa}}\ and\ \bibinfo {author} {\bibfnamefont {Y.}~\bibnamefont
  {Tokura}},\ }\href {http://dx.doi.org/10.1038/nnano.2013.243} {\bibfield
  {journal} {\bibinfo  {journal} {Nat Nano}\ }\textbf {\bibinfo {volume} {8}},\
  \bibinfo {pages} {899} (\bibinfo {year} {2013}{\natexlab{a}})}\BibitemShut
  {NoStop}%
\bibitem [{\citenamefont {Fert}\ \emph {et~al.}(2013)\citenamefont {Fert},
  \citenamefont {Cros},\ and\ \citenamefont {Sampaio}}]{Fert2013}%
  \BibitemOpen
  \bibfield  {author} {\bibinfo {author} {\bibfnamefont {A.}~\bibnamefont
  {Fert}}, \bibinfo {author} {\bibfnamefont {V.}~\bibnamefont {Cros}}, \ and\
  \bibinfo {author} {\bibfnamefont {J.}~\bibnamefont {Sampaio}},\ }\href
  {http://dx.doi.org/10.1038/nnano.2013.29} {\bibfield  {journal} {\bibinfo
  {journal} {Nat Nano}\ }\textbf {\bibinfo {volume} {8}},\ \bibinfo {pages}
  {152} (\bibinfo {year} {2013})}\BibitemShut {NoStop}%
\bibitem [{\citenamefont {Han}(2017)}]{han2017skyrmions}%
  \BibitemOpen
  \bibfield  {author} {\bibinfo {author} {\bibfnamefont {J.}~\bibnamefont
  {Han}},\ }\href {https://books.google.fi/books?id=tng5DwAAQBAJ} {\emph
  {\bibinfo {title} {Skyrmions in Condensed Matter}}},\ Springer Tracts in
  Modern Physics\ (\bibinfo  {publisher} {Springer International Publishing},\
  \bibinfo {year} {2017})\BibitemShut {NoStop}%
\bibitem [{\citenamefont {Neubauer}\ \emph {et~al.}(2009)\citenamefont
  {Neubauer}, \citenamefont {Pfleiderer}, \citenamefont {Binz}, \citenamefont
  {Rosch}, \citenamefont {Ritz}, \citenamefont {Niklowitz},\ and\ \citenamefont
  {B${\rm \ddot o}$ni}}]{Neubauer2009}%
  \BibitemOpen
  \bibfield  {author} {\bibinfo {author} {\bibfnamefont {A.}~\bibnamefont
  {Neubauer}}, \bibinfo {author} {\bibfnamefont {C.}~\bibnamefont
  {Pfleiderer}}, \bibinfo {author} {\bibfnamefont {B.}~\bibnamefont {Binz}},
  \bibinfo {author} {\bibfnamefont {A.}~\bibnamefont {Rosch}}, \bibinfo
  {author} {\bibfnamefont {R.}~\bibnamefont {Ritz}}, \bibinfo {author}
  {\bibfnamefont {P.~G.}\ \bibnamefont {Niklowitz}}, \ and\ \bibinfo {author}
  {\bibfnamefont {P.}~\bibnamefont {B${\rm \ddot o}$ni}},\ }\href
  {http://link.aps.org/doi/10.1103/PhysRevLett.102.186602} {\bibfield
  {journal} {\bibinfo  {journal} {Phys. Rev. Lett.}\ }\textbf {\bibinfo
  {volume} {102}},\ \bibinfo {pages} {186602} (\bibinfo {year}
  {2009})}\BibitemShut {NoStop}%
\bibitem [{\citenamefont {Schulz}\ \emph {et~al.}(2012)\citenamefont {Schulz},
  \citenamefont {Ritz}, \citenamefont {Bauer}, \citenamefont {Halder},
  \citenamefont {Wagner}, \citenamefont {Franz}, \citenamefont {Pfleiderer},
  \citenamefont {Everschor}, \citenamefont {Garst},\ and\ \citenamefont
  {Rosch}}]{Schulz2012}%
  \BibitemOpen
  \bibfield  {author} {\bibinfo {author} {\bibfnamefont {T.}~\bibnamefont
  {Schulz}}, \bibinfo {author} {\bibfnamefont {R.}~\bibnamefont {Ritz}},
  \bibinfo {author} {\bibfnamefont {A.}~\bibnamefont {Bauer}}, \bibinfo
  {author} {\bibfnamefont {M.}~\bibnamefont {Halder}}, \bibinfo {author}
  {\bibfnamefont {M.}~\bibnamefont {Wagner}}, \bibinfo {author} {\bibfnamefont
  {C.}~\bibnamefont {Franz}}, \bibinfo {author} {\bibfnamefont
  {C.}~\bibnamefont {Pfleiderer}}, \bibinfo {author} {\bibfnamefont
  {K.}~\bibnamefont {Everschor}}, \bibinfo {author} {\bibfnamefont
  {M.}~\bibnamefont {Garst}}, \ and\ \bibinfo {author} {\bibfnamefont
  {A.}~\bibnamefont {Rosch}},\ }\href {http://dx.doi.org/10.1038/nphys2231}
  {\bibfield  {journal} {\bibinfo  {journal} {Nat Phys}\ }\textbf {\bibinfo
  {volume} {8}},\ \bibinfo {pages} {301} (\bibinfo {year} {2012})}\BibitemShut
  {NoStop}%
\bibitem [{\citenamefont {Liang}\ \emph {et~al.}(2015)\citenamefont {Liang},
  \citenamefont {DeGrave}, \citenamefont {Stolt}, \citenamefont {Tokura},\ and\
  \citenamefont {Jin}}]{Liang2015}%
  \BibitemOpen
  \bibfield  {author} {\bibinfo {author} {\bibfnamefont {D.}~\bibnamefont
  {Liang}}, \bibinfo {author} {\bibfnamefont {J.~P.}\ \bibnamefont {DeGrave}},
  \bibinfo {author} {\bibfnamefont {M.~J.}\ \bibnamefont {Stolt}}, \bibinfo
  {author} {\bibfnamefont {Y.}~\bibnamefont {Tokura}}, \ and\ \bibinfo {author}
  {\bibfnamefont {S.}~\bibnamefont {Jin}},\ }\href
  {http://dx.doi.org/10.1038/ncomms9217} {\bibfield  {journal} {\bibinfo
  {journal} {Nature Communications}\ }\textbf {\bibinfo {volume} {6}},\
  \bibinfo {pages} {8217} (\bibinfo {year} {2015})}\BibitemShut {NoStop}%
\bibitem [{Note1()}]{Note1}%
  \BibitemOpen
  \bibinfo {note} {The chirality-sensitive terms in the free energy were
  calculated for the system consisting of the Josephson junction through
  magnetic trilayer \cite {Kulagina2014}. It has been obtained that the
  presence of scattering barriers separating ferromagnetic regions is crucial
  for such terms to be non-zero. In the present work we show that the
  chirality-selective energy arise in the generic problem with three magnetic
  impurities and no extra conditions are needed. Also we demonstrate that such
  energy contributions appear in the systems with continuous spin textures like
  magnetic spiral and skyrmion.}\BibitemShut {Stop}%
\bibitem [{\citenamefont {YU}(1965)}]{L.U.H.1965}%
  \BibitemOpen
  \bibfield  {author} {\bibinfo {author} {\bibfnamefont {L.~U.~H.}\
  \bibnamefont {YU}},\ }\href {\doibase 10.7498/aps.21.75} {\bibfield
  {journal} {\bibinfo  {journal} {Acta Physica Sinica}\ }\textbf {\bibinfo
  {volume} {21}},\ \bibinfo {pages} {75} (\bibinfo {year} {1965})}\BibitemShut
  {NoStop}%
\bibitem [{\citenamefont {Shiba}(1968)}]{Shiba1968}%
  \BibitemOpen
  \bibfield  {author} {\bibinfo {author} {\bibfnamefont {H.}~\bibnamefont
  {Shiba}},\ }\href {\doibase 10.1143/ptp.40.435} {\bibfield  {journal}
  {\bibinfo  {journal} {Progress of Theoretical Physics}\ }\textbf {\bibinfo
  {volume} {40}},\ \bibinfo {pages} {435} (\bibinfo {year} {1968})}\BibitemShut
  {NoStop}%
\bibitem [{\citenamefont {A. I.}(1969)}]{Rusinov1969}%
  \BibitemOpen
  \bibfield  {author} {\bibinfo {author} {\bibfnamefont {R.}~\bibnamefont
  {A. I.}},\ }\href@noop {} {\bibfield  {journal} {\bibinfo  {journal}
  {JETP}\ }\textbf {\bibinfo {volume} {9}},\ \bibinfo {pages} {85} (\bibinfo
  {year} {1969})},\ \bibinfo {note} {[ZhETF, {\bf 56}, 2047, (1969)
  ]}\BibitemShut {NoStop}%
\bibitem [{\citenamefont {Yao}\ \emph {et~al.}(2014)\citenamefont {Yao},
  \citenamefont {Glazman}, \citenamefont {Demler}, \citenamefont {Lukin},\ and\
  \citenamefont {Sau}}]{PhysRevLett.113.087202}%
  \BibitemOpen
  \bibfield  {author} {\bibinfo {author} {\bibfnamefont {N.~Y.}\ \bibnamefont
  {Yao}}, \bibinfo {author} {\bibfnamefont {L.~I.}\ \bibnamefont {Glazman}},
  \bibinfo {author} {\bibfnamefont {E.~A.}\ \bibnamefont {Demler}}, \bibinfo
  {author} {\bibfnamefont {M.~D.}\ \bibnamefont {Lukin}}, \ and\ \bibinfo
  {author} {\bibfnamefont {J.~D.}\ \bibnamefont {Sau}},\ }\href {\doibase
  10.1103/PhysRevLett.113.087202} {\bibfield  {journal} {\bibinfo  {journal}
  {Phys. Rev. Lett.}\ }\textbf {\bibinfo {volume} {113}},\ \bibinfo {pages}
  {087202} (\bibinfo {year} {2014})}\BibitemShut {NoStop}%
\bibitem [{\citenamefont {Fominov}\ \emph {et~al.}(2011)\citenamefont
  {Fominov}, \citenamefont {Houzet},\ and\ \citenamefont
  {Glazman}}]{PhysRevB.84.224517}%
  \BibitemOpen
  \bibfield  {author} {\bibinfo {author} {\bibfnamefont {Y.~V.}\ \bibnamefont
  {Fominov}}, \bibinfo {author} {\bibfnamefont {M.}~\bibnamefont {Houzet}}, \
  and\ \bibinfo {author} {\bibfnamefont {L.~I.}\ \bibnamefont {Glazman}},\
  }\href {\doibase 10.1103/PhysRevB.84.224517} {\bibfield  {journal} {\bibinfo
  {journal} {Phys. Rev. B}\ }\textbf {\bibinfo {volume} {84}},\ \bibinfo
  {pages} {224517} (\bibinfo {year} {2011})}\BibitemShut {NoStop}%
\bibitem [{\citenamefont {Balatsky}\ \emph {et~al.}(2006)\citenamefont
  {Balatsky}, \citenamefont {Vekhter},\ and\ \citenamefont
  {Zhu}}]{RevModPhys.78.373}%
  \BibitemOpen
  \bibfield  {author} {\bibinfo {author} {\bibfnamefont {A.~V.}\ \bibnamefont
  {Balatsky}}, \bibinfo {author} {\bibfnamefont {I.}~\bibnamefont {Vekhter}}, \
  and\ \bibinfo {author} {\bibfnamefont {J.-X.}\ \bibnamefont {Zhu}},\ }\href
  {\doibase 10.1103/RevModPhys.78.373} {\bibfield  {journal} {\bibinfo
  {journal} {Rev. Mod. Phys.}\ }\textbf {\bibinfo {volume} {78}},\ \bibinfo
  {pages} {373} (\bibinfo {year} {2006})}\BibitemShut {NoStop}%
\bibitem [{\citenamefont {Martin}\ and\ \citenamefont
  {Morpurgo}(2012)}]{PhysRevB.85.144505}%
  \BibitemOpen
  \bibfield  {author} {\bibinfo {author} {\bibfnamefont {I.}~\bibnamefont
  {Martin}}\ and\ \bibinfo {author} {\bibfnamefont {A.~F.}\ \bibnamefont
  {Morpurgo}},\ }\href {\doibase 10.1103/PhysRevB.85.144505} {\bibfield
  {journal} {\bibinfo  {journal} {Phys. Rev. B}\ }\textbf {\bibinfo {volume}
  {85}},\ \bibinfo {pages} {144505} (\bibinfo {year} {2012})}\BibitemShut
  {NoStop}%
\bibitem [{\citenamefont {Kjaergaard}\ \emph {et~al.}(2012)\citenamefont
  {Kjaergaard}, \citenamefont {W\"olms},\ and\ \citenamefont
  {Flensberg}}]{PhysRevB.85.020503}%
  \BibitemOpen
  \bibfield  {author} {\bibinfo {author} {\bibfnamefont {M.}~\bibnamefont
  {Kjaergaard}}, \bibinfo {author} {\bibfnamefont {K.}~\bibnamefont {W\"olms}},
  \ and\ \bibinfo {author} {\bibfnamefont {K.}~\bibnamefont {Flensberg}},\
  }\href {\doibase 10.1103/PhysRevB.85.020503} {\bibfield  {journal} {\bibinfo
  {journal} {Phys. Rev. B}\ }\textbf {\bibinfo {volume} {85}},\ \bibinfo
  {pages} {020503} (\bibinfo {year} {2012})}\BibitemShut {NoStop}%
\bibitem [{\citenamefont {Choy}\ \emph {et~al.}(2011)\citenamefont {Choy},
  \citenamefont {Edge}, \citenamefont {Akhmerov},\ and\ \citenamefont
  {Beenakker}}]{PhysRevB.84.195442}%
  \BibitemOpen
  \bibfield  {author} {\bibinfo {author} {\bibfnamefont {T.-P.}\ \bibnamefont
  {Choy}}, \bibinfo {author} {\bibfnamefont {J.~M.}\ \bibnamefont {Edge}},
  \bibinfo {author} {\bibfnamefont {A.~R.}\ \bibnamefont {Akhmerov}}, \ and\
  \bibinfo {author} {\bibfnamefont {C.~W.~J.}\ \bibnamefont {Beenakker}},\
  }\href {\doibase 10.1103/PhysRevB.84.195442} {\bibfield  {journal} {\bibinfo
  {journal} {Phys. Rev. B}\ }\textbf {\bibinfo {volume} {84}},\ \bibinfo
  {pages} {195442} (\bibinfo {year} {2011})}\BibitemShut {NoStop}%
\bibitem [{\citenamefont {P\"oyh\"onen}\ \emph {et~al.}(2014)\citenamefont
  {P\"oyh\"onen}, \citenamefont {Weststr\"om}, \citenamefont {R\"ontynen},\
  and\ \citenamefont {Ojanen}}]{PhysRevB.89.115109}%
  \BibitemOpen
  \bibfield  {author} {\bibinfo {author} {\bibfnamefont {K.}~\bibnamefont
  {P\"oyh\"onen}}, \bibinfo {author} {\bibfnamefont {A.}~\bibnamefont
  {Weststr\"om}}, \bibinfo {author} {\bibfnamefont {J.}~\bibnamefont
  {R\"ontynen}}, \ and\ \bibinfo {author} {\bibfnamefont {T.}~\bibnamefont
  {Ojanen}},\ }\href {\doibase 10.1103/PhysRevB.89.115109} {\bibfield
  {journal} {\bibinfo  {journal} {Phys. Rev. B}\ }\textbf {\bibinfo {volume}
  {89}},\ \bibinfo {pages} {115109} (\bibinfo {year} {2014})}\BibitemShut
  {NoStop}%
\bibitem [{\citenamefont {Vazifeh}\ and\ \citenamefont
  {Franz}(2013)}]{PhysRevLett.111.206802}%
  \BibitemOpen
  \bibfield  {author} {\bibinfo {author} {\bibfnamefont {M.~M.}\ \bibnamefont
  {Vazifeh}}\ and\ \bibinfo {author} {\bibfnamefont {M.}~\bibnamefont
  {Franz}},\ }\href {\doibase 10.1103/PhysRevLett.111.206802} {\bibfield
  {journal} {\bibinfo  {journal} {Phys. Rev. Lett.}\ }\textbf {\bibinfo
  {volume} {111}},\ \bibinfo {pages} {206802} (\bibinfo {year}
  {2013})}\BibitemShut {NoStop}%
\bibitem [{\citenamefont {Braunecker}\ and\ \citenamefont
  {Simon}(2013)}]{PhysRevLett.111.147202}%
  \BibitemOpen
  \bibfield  {author} {\bibinfo {author} {\bibfnamefont {B.}~\bibnamefont
  {Braunecker}}\ and\ \bibinfo {author} {\bibfnamefont {P.}~\bibnamefont
  {Simon}},\ }\href {\doibase 10.1103/PhysRevLett.111.147202} {\bibfield
  {journal} {\bibinfo  {journal} {Phys. Rev. Lett.}\ }\textbf {\bibinfo
  {volume} {111}},\ \bibinfo {pages} {147202} (\bibinfo {year}
  {2013})}\BibitemShut {NoStop}%
\bibitem [{\citenamefont {Klinovaja}\ \emph {et~al.}(2013)\citenamefont
  {Klinovaja}, \citenamefont {Stano}, \citenamefont {Yazdani},\ and\
  \citenamefont {Loss}}]{PhysRevLett.111.186805}%
  \BibitemOpen
  \bibfield  {author} {\bibinfo {author} {\bibfnamefont {J.}~\bibnamefont
  {Klinovaja}}, \bibinfo {author} {\bibfnamefont {P.}~\bibnamefont {Stano}},
  \bibinfo {author} {\bibfnamefont {A.}~\bibnamefont {Yazdani}}, \ and\
  \bibinfo {author} {\bibfnamefont {D.}~\bibnamefont {Loss}},\ }\href {\doibase
  10.1103/PhysRevLett.111.186805} {\bibfield  {journal} {\bibinfo  {journal}
  {Phys. Rev. Lett.}\ }\textbf {\bibinfo {volume} {111}},\ \bibinfo {pages}
  {186805} (\bibinfo {year} {2013})}\BibitemShut {NoStop}%
\bibitem [{\citenamefont {Nakosai}\ \emph {et~al.}(2013)\citenamefont
  {Nakosai}, \citenamefont {Tanaka},\ and\ \citenamefont
  {Nagaosa}}]{PhysRevB.88.180503}%
  \BibitemOpen
  \bibfield  {author} {\bibinfo {author} {\bibfnamefont {S.}~\bibnamefont
  {Nakosai}}, \bibinfo {author} {\bibfnamefont {Y.}~\bibnamefont {Tanaka}}, \
  and\ \bibinfo {author} {\bibfnamefont {N.}~\bibnamefont {Nagaosa}},\ }\href
  {\doibase 10.1103/PhysRevB.88.180503} {\bibfield  {journal} {\bibinfo
  {journal} {Phys. Rev. B}\ }\textbf {\bibinfo {volume} {88}},\ \bibinfo
  {pages} {180503} (\bibinfo {year} {2013})}\BibitemShut {NoStop}%
\bibitem [{\citenamefont {Nadj-Perge}\ \emph {et~al.}(2013)\citenamefont
  {Nadj-Perge}, \citenamefont {Drozdov}, \citenamefont {Bernevig},\ and\
  \citenamefont {Yazdani}}]{PhysRevB.88.020407}%
  \BibitemOpen
  \bibfield  {author} {\bibinfo {author} {\bibfnamefont {S.}~\bibnamefont
  {Nadj-Perge}}, \bibinfo {author} {\bibfnamefont {I.~K.}\ \bibnamefont
  {Drozdov}}, \bibinfo {author} {\bibfnamefont {B.~A.}\ \bibnamefont
  {Bernevig}}, \ and\ \bibinfo {author} {\bibfnamefont {A.}~\bibnamefont
  {Yazdani}},\ }\href {\doibase 10.1103/PhysRevB.88.020407} {\bibfield
  {journal} {\bibinfo  {journal} {Phys. Rev. B}\ }\textbf {\bibinfo {volume}
  {88}},\ \bibinfo {pages} {020407} (\bibinfo {year} {2013})}\BibitemShut
  {NoStop}%
\bibitem [{\citenamefont {Pientka}\ \emph {et~al.}(2014)\citenamefont
  {Pientka}, \citenamefont {Glazman},\ and\ \citenamefont {von
  Oppen}}]{PhysRevB.89.180505}%
  \BibitemOpen
  \bibfield  {author} {\bibinfo {author} {\bibfnamefont {F.}~\bibnamefont
  {Pientka}}, \bibinfo {author} {\bibfnamefont {L.~I.}\ \bibnamefont
  {Glazman}}, \ and\ \bibinfo {author} {\bibfnamefont {F.}~\bibnamefont {von
  Oppen}},\ }\href {\doibase 10.1103/PhysRevB.89.180505} {\bibfield  {journal}
  {\bibinfo  {journal} {Phys. Rev. B}\ }\textbf {\bibinfo {volume} {89}},\
  \bibinfo {pages} {180505} (\bibinfo {year} {2014})}\BibitemShut {NoStop}%
\bibitem [{\citenamefont {Pientka}\ \emph {et~al.}(2013)\citenamefont
  {Pientka}, \citenamefont {Glazman},\ and\ \citenamefont {von
  Oppen}}]{PhysRevB.88.155420}%
  \BibitemOpen
  \bibfield  {author} {\bibinfo {author} {\bibfnamefont {F.}~\bibnamefont
  {Pientka}}, \bibinfo {author} {\bibfnamefont {L.~I.}\ \bibnamefont
  {Glazman}}, \ and\ \bibinfo {author} {\bibfnamefont {F.}~\bibnamefont {von
  Oppen}},\ }\href {\doibase 10.1103/PhysRevB.88.155420} {\bibfield  {journal}
  {\bibinfo  {journal} {Phys. Rev. B}\ }\textbf {\bibinfo {volume} {88}},\
  \bibinfo {pages} {155420} (\bibinfo {year} {2013})}\BibitemShut {NoStop}%
\bibitem [{\citenamefont {Yazdani}\ \emph {et~al.}(1997)\citenamefont
  {Yazdani}, \citenamefont {Jones}, \citenamefont {Lutz}, \citenamefont
  {Crommie},\ and\ \citenamefont {Eigler}}]{Yazdani1767}%
  \BibitemOpen
  \bibfield  {author} {\bibinfo {author} {\bibfnamefont {A.}~\bibnamefont
  {Yazdani}}, \bibinfo {author} {\bibfnamefont {B.~A.}\ \bibnamefont {Jones}},
  \bibinfo {author} {\bibfnamefont {C.~P.}\ \bibnamefont {Lutz}}, \bibinfo
  {author} {\bibfnamefont {M.~F.}\ \bibnamefont {Crommie}}, \ and\ \bibinfo
  {author} {\bibfnamefont {D.~M.}\ \bibnamefont {Eigler}},\ }\href {\doibase
  10.1126/science.275.5307.1767} {\bibfield  {journal} {\bibinfo  {journal}
  {Science}\ }\textbf {\bibinfo {volume} {275}},\ \bibinfo {pages} {1767}
  (\bibinfo {year} {1997})},\ \Eprint
  {http://arxiv.org/abs/http://science.sciencemag.org/content/275/5307/1767.full.pdf}
  {http://science.sciencemag.org/content/275/5307/1767.full.pdf} \BibitemShut
  {NoStop}%
\bibitem [{\citenamefont {Kulagina}\ and\ \citenamefont
  {Linder}(2014)}]{Kulagina2014}%
  \BibitemOpen
  \bibfield  {author} {\bibinfo {author} {\bibfnamefont {I.}~\bibnamefont
  {Kulagina}}\ and\ \bibinfo {author} {\bibfnamefont {J.}~\bibnamefont
  {Linder}},\ }\href {http://link.aps.org/doi/10.1103/PhysRevB.90.054504}
  {\bibfield  {journal} {\bibinfo  {journal} {Phys. Rev. B}\ }\textbf {\bibinfo
  {volume} {90}},\ \bibinfo {pages} {054504} (\bibinfo {year}
  {2014})}\BibitemShut {NoStop}%
\bibitem [{\citenamefont {Bobkova}\ and\ \citenamefont
  {Barash}(2004)}]{Bobkova2004}%
  \BibitemOpen
  \bibfield  {author} {\bibinfo {author} {\bibfnamefont {I.~V.}\ \bibnamefont
  {Bobkova}}\ and\ \bibinfo {author} {\bibfnamefont {Y.~S.}\ \bibnamefont
  {Barash}},\ }\href {https://doi.org/10.1134/1.1839298} {\bibfield  {journal}
  {\bibinfo  {journal} {Journal of Experimental and Theoretical Physics
  Letters}\ }\textbf {\bibinfo {volume} {80}},\ \bibinfo {pages} {494}
  (\bibinfo {year} {2004})}\BibitemShut {NoStop}%
\bibitem [{\citenamefont {Dolcini}\ \emph {et~al.}(2015)\citenamefont
  {Dolcini}, \citenamefont {Houzet},\ and\ \citenamefont
  {Meyer}}]{Dolcini2015}%
  \BibitemOpen
  \bibfield  {author} {\bibinfo {author} {\bibfnamefont {F.}~\bibnamefont
  {Dolcini}}, \bibinfo {author} {\bibfnamefont {M.}~\bibnamefont {Houzet}}, \
  and\ \bibinfo {author} {\bibfnamefont {J.~S.}\ \bibnamefont {Meyer}},\ }\href
  {https://link.aps.org/doi/10.1103/PhysRevB.92.035428} {\bibfield  {journal}
  {\bibinfo  {journal} {Phys. Rev. B}\ }\textbf {\bibinfo {volume} {92}},\
  \bibinfo {pages} {035428} (\bibinfo {year} {2015})}\BibitemShut {NoStop}%
\bibitem [{\citenamefont {Pershoguba}\ \emph {et~al.}(2015)\citenamefont
  {Pershoguba}, \citenamefont {Bjornson}, \citenamefont {Black-Schaffer},\ and\
  \citenamefont {Balatsky}}]{Pershoguba2015}%
  \BibitemOpen
  \bibfield  {author} {\bibinfo {author} {\bibfnamefont {S.~S.}\ \bibnamefont
  {Pershoguba}}, \bibinfo {author} {\bibfnamefont {K.}~\bibnamefont
  {Bjornson}}, \bibinfo {author} {\bibfnamefont {A.~M.}\ \bibnamefont
  {Black-Schaffer}}, \ and\ \bibinfo {author} {\bibfnamefont {A.~V.}\
  \bibnamefont {Balatsky}},\ }\href
  {https://doi.org/10.1103/PhysRevLett.115.116602} {\bibfield  {journal}
  {\bibinfo  {journal} {Phys. Rev. Lett.}\ }\textbf {\bibinfo {volume} {115}},\
  \bibinfo {pages} {116602} (\bibinfo {year} {2015})}\BibitemShut {NoStop}%
\bibitem [{\citenamefont {Mal'shukov}(2016)}]{Malshukov2016}%
  \BibitemOpen
  \bibfield  {author} {\bibinfo {author} {\bibfnamefont {A.~G.}\ \bibnamefont
  {Mal'shukov}},\ }\href {https://doi.org/10.1103/PhysRevB.93.054511}
  {\bibfield  {journal} {\bibinfo  {journal} {Phys. Rev. B}\ }\textbf {\bibinfo
  {volume} {93}},\ \bibinfo {pages} {054511} (\bibinfo {year}
  {2016})}\BibitemShut {NoStop}%
\bibitem [{\citenamefont {Mironov}\ and\ \citenamefont
  {Buzdin}(2017)}]{Mironov2017}%
  \BibitemOpen
  \bibfield  {author} {\bibinfo {author} {\bibfnamefont {S.}~\bibnamefont
  {Mironov}}\ and\ \bibinfo {author} {\bibfnamefont {A.}~\bibnamefont
  {Buzdin}},\ }\href {http://link.aps.org/doi/10.1103/PhysRevLett.118.077001}
  {\bibfield  {journal} {\bibinfo  {journal} {Phys. Rev. Lett.}\ }\textbf
  {\bibinfo {volume} {118}},\ \bibinfo {pages} {077001} (\bibinfo {year}
  {2017})}\BibitemShut {NoStop}%
\bibitem [{\citenamefont {Edel'shtein}(1989)}]{edelshtein1989}%
  \BibitemOpen
  \bibfield  {author} {\bibinfo {author} {\bibfnamefont {V.}~\bibnamefont
  {Edel'shtein}},\ }\href@noop {} {\bibfield  {journal} {\bibinfo  {journal}
  {Sov. Phys. JETP}\ }\textbf {\bibinfo {volume} {68}},\ \bibinfo {pages}
  {1244} (\bibinfo {year} {1989})}\BibitemShut {NoStop}%
\bibitem [{\citenamefont {Barzykin}\ and\ \citenamefont
  {Gor’kov}(2002)}]{Barzykin2002}%
  \BibitemOpen
  \bibfield  {author} {\bibinfo {author} {\bibfnamefont {V.}~\bibnamefont
  {Barzykin}}\ and\ \bibinfo {author} {\bibfnamefont {L.~P.}\ \bibnamefont
  {Gor’kov}},\ }\href {https://doi.org/10.1103/PhysRevLett.89.227002}
  {\bibfield  {journal} {\bibinfo  {journal} {Phys. Rev. Lett.}\ }\textbf
  {\bibinfo {volume} {89}},\ \bibinfo {pages} {227002} (\bibinfo {year}
  {2002})}\BibitemShut {NoStop}%
\bibitem [{\citenamefont {Samokhin}(2004)}]{PhysRevB.70.104521}%
  \BibitemOpen
  \bibfield  {author} {\bibinfo {author} {\bibfnamefont {K.~V.}\ \bibnamefont
  {Samokhin}},\ }\href {\doibase 10.1103/PhysRevB.70.104521} {\bibfield
  {journal} {\bibinfo  {journal} {Phys. Rev. B}\ }\textbf {\bibinfo {volume}
  {70}},\ \bibinfo {pages} {104521} (\bibinfo {year} {2004})}\BibitemShut
  {NoStop}%
\bibitem [{\citenamefont {Kaur}\ \emph {et~al.}(2005)\citenamefont {Kaur},
  \citenamefont {Agterberg},\ and\ \citenamefont
  {Sigrist}}]{PhysRevLett.94.137002}%
  \BibitemOpen
  \bibfield  {author} {\bibinfo {author} {\bibfnamefont {R.~P.}\ \bibnamefont
  {Kaur}}, \bibinfo {author} {\bibfnamefont {D.~F.}\ \bibnamefont {Agterberg}},
  \ and\ \bibinfo {author} {\bibfnamefont {M.}~\bibnamefont {Sigrist}},\ }\href
  {\doibase 10.1103/PhysRevLett.94.137002} {\bibfield  {journal} {\bibinfo
  {journal} {Phys. Rev. Lett.}\ }\textbf {\bibinfo {volume} {94}},\ \bibinfo
  {pages} {137002} (\bibinfo {year} {2005})}\BibitemShut {NoStop}%
\bibitem [{\citenamefont {Dimitrova}\ and\ \citenamefont
  {Feigel'man}(2007)}]{PhysRevB.76.014522}%
  \BibitemOpen
  \bibfield  {author} {\bibinfo {author} {\bibfnamefont {O.}~\bibnamefont
  {Dimitrova}}\ and\ \bibinfo {author} {\bibfnamefont {M.~V.}\ \bibnamefont
  {Feigel'man}},\ }\href {\doibase 10.1103/PhysRevB.76.014522} {\bibfield
  {journal} {\bibinfo  {journal} {Phys. Rev. B}\ }\textbf {\bibinfo {volume}
  {76}},\ \bibinfo {pages} {014522} (\bibinfo {year} {2007})}\BibitemShut
  {NoStop}%
\bibitem [{\citenamefont {Houzet}\ and\ \citenamefont
  {Meyer}(2015)}]{PhysRevB.92.014509}%
  \BibitemOpen
  \bibfield  {author} {\bibinfo {author} {\bibfnamefont {M.}~\bibnamefont
  {Houzet}}\ and\ \bibinfo {author} {\bibfnamefont {J.~S.}\ \bibnamefont
  {Meyer}},\ }\href {\doibase 10.1103/PhysRevB.92.014509} {\bibfield  {journal}
  {\bibinfo  {journal} {Phys. Rev. B}\ }\textbf {\bibinfo {volume} {92}},\
  \bibinfo {pages} {014509} (\bibinfo {year} {2015})}\BibitemShut {NoStop}%
\bibitem [{\citenamefont {Krive}\ \emph {et~al.}(2004)\citenamefont {Krive},
  \citenamefont {Gorelik}, \citenamefont {Shekhter},\ and\ \citenamefont
  {Jonson}}]{krive04}%
  \BibitemOpen
  \bibfield  {author} {\bibinfo {author} {\bibfnamefont {I.}~\bibnamefont
  {Krive}}, \bibinfo {author} {\bibfnamefont {L.}~\bibnamefont {Gorelik}},
  \bibinfo {author} {\bibfnamefont {R.}~\bibnamefont {Shekhter}}, \ and\
  \bibinfo {author} {\bibfnamefont {M.}~\bibnamefont {Jonson}},\ }\href@noop {}
  {\bibfield  {journal} {\bibinfo  {journal} {Phys. Nizk. Temp.}\ }\textbf
  {\bibinfo {volume} {30}},\ \bibinfo {pages} {535} (\bibinfo {year}
  {2004})}\BibitemShut {NoStop}%
\bibitem [{\citenamefont {Braude}\ and\ \citenamefont
  {Nazarov}(2007)}]{Braude2007}%
  \BibitemOpen
  \bibfield  {author} {\bibinfo {author} {\bibfnamefont {V.}~\bibnamefont
  {Braude}}\ and\ \bibinfo {author} {\bibfnamefont {Y.~V.}\ \bibnamefont
  {Nazarov}},\ }\href {http://link.aps.org/doi/10.1103/PhysRevLett.98.077003}
  {\bibfield  {journal} {\bibinfo  {journal} {Phys. Rev. Lett.}\ }\textbf
  {\bibinfo {volume} {98}},\ \bibinfo {pages} {077003} (\bibinfo {year}
  {2007})}\BibitemShut {NoStop}%
\bibitem [{\citenamefont {Asano}\ \emph {et~al.}(2007)\citenamefont {Asano},
  \citenamefont {Sawa}, \citenamefont {Tanaka},\ and\ \citenamefont
  {Golubov}}]{Asano2007}%
  \BibitemOpen
  \bibfield  {author} {\bibinfo {author} {\bibfnamefont {Y.}~\bibnamefont
  {Asano}}, \bibinfo {author} {\bibfnamefont {Y.}~\bibnamefont {Sawa}},
  \bibinfo {author} {\bibfnamefont {Y.}~\bibnamefont {Tanaka}}, \ and\ \bibinfo
  {author} {\bibfnamefont {A.~A.}\ \bibnamefont {Golubov}},\ }\href
  {https://link.aps.org/doi/10.1103/PhysRevB.76.224525} {\bibfield  {journal}
  {\bibinfo  {journal} {Phys. Rev. B}\ }\textbf {\bibinfo {volume} {76}},\
  \bibinfo {pages} {224525} (\bibinfo {year} {2007})}\BibitemShut {NoStop}%
\bibitem [{\citenamefont {Reynoso}\ \emph {et~al.}(2008)\citenamefont
  {Reynoso}, \citenamefont {Usaj}, \citenamefont {Balseiro}, \citenamefont
  {Feinberg},\ and\ \citenamefont {Avignon}}]{Reynoso2008}%
  \BibitemOpen
  \bibfield  {author} {\bibinfo {author} {\bibfnamefont {A.~A.}\ \bibnamefont
  {Reynoso}}, \bibinfo {author} {\bibfnamefont {G.}~\bibnamefont {Usaj}},
  \bibinfo {author} {\bibfnamefont {C.~A.}\ \bibnamefont {Balseiro}}, \bibinfo
  {author} {\bibfnamefont {D.}~\bibnamefont {Feinberg}}, \ and\ \bibinfo
  {author} {\bibfnamefont {M.}~\bibnamefont {Avignon}},\ }\href
  {http://link.aps.org/doi/10.1103/PhysRevLett.101.107001} {\bibfield
  {journal} {\bibinfo  {journal} {Phys. Rev. Lett.}\ }\textbf {\bibinfo
  {volume} {101}},\ \bibinfo {pages} {107001} (\bibinfo {year}
  {2008})}\BibitemShut {NoStop}%
\bibitem [{\citenamefont {Eschrig}\ and\ \citenamefont
  {Lofwander}(2008)}]{Eschrig2008}%
  \BibitemOpen
  \bibfield  {author} {\bibinfo {author} {\bibfnamefont {M.}~\bibnamefont
  {Eschrig}}\ and\ \bibinfo {author} {\bibfnamefont {T.}~\bibnamefont
  {Lofwander}},\ }\href {http://dx.doi.org/10.1038/nphys831} {\bibfield
  {journal} {\bibinfo  {journal} {Nat Phys}\ }\textbf {\bibinfo {volume} {4}},\
  \bibinfo {pages} {138} (\bibinfo {year} {2008})}\BibitemShut {NoStop}%
\bibitem [{\citenamefont {Buzdin}(2008)}]{PhysRevLett.101.107005}%
  \BibitemOpen
  \bibfield  {author} {\bibinfo {author} {\bibfnamefont {A.}~\bibnamefont
  {Buzdin}},\ }\href {https://link.aps.org/doi/10.1103/PhysRevLett.101.107005}
  {\bibfield  {journal} {\bibinfo  {journal} {Phys. Rev. Lett.}\ }\textbf
  {\bibinfo {volume} {101}},\ \bibinfo {pages} {107005} (\bibinfo {year}
  {2008})}\BibitemShut {NoStop}%
\bibitem [{\citenamefont {Tanaka}\ \emph {et~al.}(2009)\citenamefont {Tanaka},
  \citenamefont {Yokoyama},\ and\ \citenamefont {Nagaosa}}]{Tanaka2009}%
  \BibitemOpen
  \bibfield  {author} {\bibinfo {author} {\bibfnamefont {Y.}~\bibnamefont
  {Tanaka}}, \bibinfo {author} {\bibfnamefont {T.}~\bibnamefont {Yokoyama}}, \
  and\ \bibinfo {author} {\bibfnamefont {N.}~\bibnamefont {Nagaosa}},\ }\href
  {https://link.aps.org/doi/10.1103/PhysRevLett.103.107002} {\bibfield
  {journal} {\bibinfo  {journal} {Phys. Rev. Lett.}\ }\textbf {\bibinfo
  {volume} {103}},\ \bibinfo {pages} {107002} (\bibinfo {year}
  {2009})}\BibitemShut {NoStop}%
\bibitem [{\citenamefont {Grein}\ \emph {et~al.}(2009)\citenamefont {Grein},
  \citenamefont {Eschrig}, \citenamefont {Metalidis},\ and\ \citenamefont
  {Sch${\rm o}$n}}]{Grein2009}%
  \BibitemOpen
  \bibfield  {author} {\bibinfo {author} {\bibfnamefont {R.}~\bibnamefont
  {Grein}}, \bibinfo {author} {\bibfnamefont {M.}~\bibnamefont {Eschrig}},
  \bibinfo {author} {\bibfnamefont {G.}~\bibnamefont {Metalidis}}, \ and\
  \bibinfo {author} {\bibfnamefont {G.}~\bibnamefont {Sch${\rm o}$n}},\ }\href
  {http://link.aps.org/doi/10.1103/PhysRevLett.102.227005} {\bibfield
  {journal} {\bibinfo  {journal} {Phys. Rev. Lett.}\ }\textbf {\bibinfo
  {volume} {102}},\ \bibinfo {pages} {227005} (\bibinfo {year}
  {2009})}\BibitemShut {NoStop}%
\bibitem [{\citenamefont {Zazunov}\ \emph {et~al.}(2009)\citenamefont
  {Zazunov}, \citenamefont {Egger}, \citenamefont {Jonckheere},\ and\
  \citenamefont {Martin}}]{Zazunov2009}%
  \BibitemOpen
  \bibfield  {author} {\bibinfo {author} {\bibfnamefont {A.}~\bibnamefont
  {Zazunov}}, \bibinfo {author} {\bibfnamefont {R.}~\bibnamefont {Egger}},
  \bibinfo {author} {\bibfnamefont {T.}~\bibnamefont {Jonckheere}}, \ and\
  \bibinfo {author} {\bibfnamefont {T.}~\bibnamefont {Martin}},\ }\href
  {http://link.aps.org/doi/10.1103/PhysRevLett.103.147004} {\bibfield
  {journal} {\bibinfo  {journal} {Phys. Rev. Lett.}\ }\textbf {\bibinfo
  {volume} {103}},\ \bibinfo {pages} {147004} (\bibinfo {year}
  {2009})}\BibitemShut {NoStop}%
\bibitem [{\citenamefont {Liu}\ and\ \citenamefont {Chan}(2010)}]{Liu2010}%
  \BibitemOpen
  \bibfield  {author} {\bibinfo {author} {\bibfnamefont {J.-F.}\ \bibnamefont
  {Liu}}\ and\ \bibinfo {author} {\bibfnamefont {K.~S.}\ \bibnamefont {Chan}},\
  }\href {http://link.aps.org/doi/10.1103/PhysRevB.82.184533} {\bibfield
  {journal} {\bibinfo  {journal} {Phys. Rev. B}\ }\textbf {\bibinfo {volume}
  {82}},\ \bibinfo {pages} {184533} (\bibinfo {year} {2010})}\BibitemShut
  {NoStop}%
\bibitem [{\citenamefont {Malshukov}\ \emph {et~al.}(2010)\citenamefont
  {Malshukov}, \citenamefont {Sadjina},\ and\ \citenamefont
  {Brataas}}]{Malshukov2010}%
  \BibitemOpen
  \bibfield  {author} {\bibinfo {author} {\bibfnamefont {A.~G.}\ \bibnamefont
  {Malshukov}}, \bibinfo {author} {\bibfnamefont {S.}~\bibnamefont {Sadjina}},
  \ and\ \bibinfo {author} {\bibfnamefont {A.}~\bibnamefont {Brataas}},\ }\href
  {https://doi.org/10.1103/PhysRevB.81.060502} {\bibfield  {journal} {\bibinfo
  {journal} {Phys. Rev. B}\ }\textbf {\bibinfo {volume} {81}},\ \bibinfo
  {pages} {060502} (\bibinfo {year} {2010})}\BibitemShut {NoStop}%
\bibitem [{\citenamefont {Zyuzin}\ \emph
  {et~al.}(2016{\natexlab{a}})\citenamefont {Zyuzin}, \citenamefont
  {Alidoust},\ and\ \citenamefont {Loss}}]{Alidoust2016}%
  \BibitemOpen
  \bibfield  {author} {\bibinfo {author} {\bibfnamefont {A.}~\bibnamefont
  {Zyuzin}}, \bibinfo {author} {\bibfnamefont {M.}~\bibnamefont {Alidoust}}, \
  and\ \bibinfo {author} {\bibfnamefont {D.}~\bibnamefont {Loss}},\ }\href
  {\doibase https://doi.org/10.1103/PhysRevB.93.214502} {\bibfield  {journal}
  {\bibinfo  {journal} {Phys. Rev. B}\ }\textbf {\bibinfo {volume} {93}},\
  \bibinfo {pages} {214502} (\bibinfo {year} {2016}{\natexlab{a}})}\BibitemShut
  {NoStop}%
\bibitem [{\citenamefont {Brunetti}\ \emph {et~al.}(2013)\citenamefont
  {Brunetti}, \citenamefont {Zazunov}, \citenamefont {Kundu},\ and\
  \citenamefont {Egger}}]{Brunetti2013}%
  \BibitemOpen
  \bibfield  {author} {\bibinfo {author} {\bibfnamefont {A.}~\bibnamefont
  {Brunetti}}, \bibinfo {author} {\bibfnamefont {A.}~\bibnamefont {Zazunov}},
  \bibinfo {author} {\bibfnamefont {A.}~\bibnamefont {Kundu}}, \ and\ \bibinfo
  {author} {\bibfnamefont {R.}~\bibnamefont {Egger}},\ }\href
  {http://link.aps.org/doi/10.1103/PhysRevB.88.144515} {\bibfield  {journal}
  {\bibinfo  {journal} {Phys. Rev. B}\ }\textbf {\bibinfo {volume} {88}},\
  \bibinfo {pages} {144515} (\bibinfo {year} {2013})}\BibitemShut {NoStop}%
\bibitem [{\citenamefont {Yokoyama}\ \emph {et~al.}(2014)\citenamefont
  {Yokoyama}, \citenamefont {Eto},\ and\ \citenamefont
  {Nazarov}}]{Yokoyama2014}%
  \BibitemOpen
  \bibfield  {author} {\bibinfo {author} {\bibfnamefont {T.}~\bibnamefont
  {Yokoyama}}, \bibinfo {author} {\bibfnamefont {M.}~\bibnamefont {Eto}}, \
  and\ \bibinfo {author} {\bibfnamefont {Y.~V.}\ \bibnamefont {Nazarov}},\
  }\href {http://link.aps.org/doi/10.1103/PhysRevB.89.195407} {\bibfield
  {journal} {\bibinfo  {journal} {Phys. Rev. B}\ }\textbf {\bibinfo {volume}
  {89}},\ \bibinfo {pages} {195407} (\bibinfo {year} {2014})}\BibitemShut
  {NoStop}%
\bibitem [{\citenamefont {Moor}\ \emph
  {et~al.}(2015{\natexlab{a}})\citenamefont {Moor}, \citenamefont {Volkov},\
  and\ \citenamefont {Efetov}}]{Moor2015}%
  \BibitemOpen
  \bibfield  {author} {\bibinfo {author} {\bibfnamefont {A.}~\bibnamefont
  {Moor}}, \bibinfo {author} {\bibfnamefont {A.~F.}\ \bibnamefont {Volkov}}, \
  and\ \bibinfo {author} {\bibfnamefont {K.~B.}\ \bibnamefont {Efetov}},\
  }\href {http://link.aps.org/doi/10.1103/PhysRevB.92.214510} {\bibfield
  {journal} {\bibinfo  {journal} {Phys. Rev. B}\ }\textbf {\bibinfo {volume}
  {92}},\ \bibinfo {pages} {214510} (\bibinfo {year}
  {2015}{\natexlab{a}})}\BibitemShut {NoStop}%
\bibitem [{\citenamefont {Moor}\ \emph
  {et~al.}(2015{\natexlab{b}})\citenamefont {Moor}, \citenamefont {Volkov},\
  and\ \citenamefont {Efetov}}]{Moor2015a}%
  \BibitemOpen
  \bibfield  {author} {\bibinfo {author} {\bibfnamefont {A.}~\bibnamefont
  {Moor}}, \bibinfo {author} {\bibfnamefont {A.~F.}\ \bibnamefont {Volkov}}, \
  and\ \bibinfo {author} {\bibfnamefont {K.~B.}\ \bibnamefont {Efetov}},\
  }\href {http://link.aps.org/doi/10.1103/PhysRevB.92.180506} {\bibfield
  {journal} {\bibinfo  {journal} {Phys. Rev. B}\ }\textbf {\bibinfo {volume}
  {92}},\ \bibinfo {pages} {180506} (\bibinfo {year}
  {2015}{\natexlab{b}})}\BibitemShut {NoStop}%
\bibitem [{\citenamefont {Bergeret}\ and\ \citenamefont
  {Tokatly}(2015)}]{Bergeret2015}%
  \BibitemOpen
  \bibfield  {author} {\bibinfo {author} {\bibfnamefont {F.~S.}\ \bibnamefont
  {Bergeret}}\ and\ \bibinfo {author} {\bibfnamefont {I.~V.}\ \bibnamefont
  {Tokatly}},\ }\href {http://stacks.iop.org/0295-5075/110/i=5/a=57005}
  {\bibfield  {journal} {\bibinfo  {journal} {EPL (Europhysics Letters)}\
  }\textbf {\bibinfo {volume} {110}},\ \bibinfo {pages} {57005} (\bibinfo
  {year} {2015})}\BibitemShut {NoStop}%
\bibitem [{\citenamefont {Campagnano}\ \emph {et~al.}(2015)\citenamefont
  {Campagnano}, \citenamefont {Lucignano}, \citenamefont {Giuliano},\ and\
  \citenamefont {Tagliacozzo}}]{Campagnano2015}%
  \BibitemOpen
  \bibfield  {author} {\bibinfo {author} {\bibfnamefont {G.}~\bibnamefont
  {Campagnano}}, \bibinfo {author} {\bibfnamefont {P.}~\bibnamefont
  {Lucignano}}, \bibinfo {author} {\bibfnamefont {D.}~\bibnamefont {Giuliano}},
  \ and\ \bibinfo {author} {\bibfnamefont {A.}~\bibnamefont {Tagliacozzo}},\
  }\href {http://stacks.iop.org/0953-8984/27/i=20/a=205301} {\bibfield
  {journal} {\bibinfo  {journal} {Journal of Physics: Condensed Matter}\
  }\textbf {\bibinfo {volume} {27}},\ \bibinfo {pages} {205301} (\bibinfo
  {year} {2015})}\BibitemShut {NoStop}%
\bibitem [{\citenamefont {Mironov}\ and\ \citenamefont
  {Buzdin}(2015)}]{Mironov2015}%
  \BibitemOpen
  \bibfield  {author} {\bibinfo {author} {\bibfnamefont {S.}~\bibnamefont
  {Mironov}}\ and\ \bibinfo {author} {\bibfnamefont {A.}~\bibnamefont
  {Buzdin}},\ }\href {http://link.aps.org/doi/10.1103/PhysRevB.92.184506}
  {\bibfield  {journal} {\bibinfo  {journal} {Phys. Rev. B}\ }\textbf {\bibinfo
  {volume} {92}},\ \bibinfo {pages} {184506} (\bibinfo {year}
  {2015})}\BibitemShut {NoStop}%
\bibitem [{\citenamefont {Konschelle}\ \emph {et~al.}(2015)\citenamefont
  {Konschelle}, \citenamefont {Tokatly},\ and\ \citenamefont
  {Bergeret}}]{PhysRevB.92.125443}%
  \BibitemOpen
  \bibfield  {author} {\bibinfo {author} {\bibfnamefont {F.}~\bibnamefont
  {Konschelle}}, \bibinfo {author} {\bibfnamefont {I.~V.}\ \bibnamefont
  {Tokatly}}, \ and\ \bibinfo {author} {\bibfnamefont {F.~S.}\ \bibnamefont
  {Bergeret}},\ }\href {\doibase 10.1103/PhysRevB.92.125443} {\bibfield
  {journal} {\bibinfo  {journal} {Phys. Rev. B}\ }\textbf {\bibinfo {volume}
  {92}},\ \bibinfo {pages} {125443} (\bibinfo {year} {2015})}\BibitemShut
  {NoStop}%
\bibitem [{\citenamefont {Kuzmanovski}\ \emph {et~al.}(2016)\citenamefont
  {Kuzmanovski}, \citenamefont {Linder},\ and\ \citenamefont
  {Black-Schaffer}}]{Kuzmanovski2016}%
  \BibitemOpen
  \bibfield  {author} {\bibinfo {author} {\bibfnamefont {D.}~\bibnamefont
  {Kuzmanovski}}, \bibinfo {author} {\bibfnamefont {J.}~\bibnamefont {Linder}},
  \ and\ \bibinfo {author} {\bibfnamefont {A.}~\bibnamefont {Black-Schaffer}},\
  }\href {https://link.aps.org/doi/10.1103/PhysRevB.94.180505} {\bibfield
  {journal} {\bibinfo  {journal} {Phys. Rev. B}\ }\textbf {\bibinfo {volume}
  {94}},\ \bibinfo {pages} {180505} (\bibinfo {year} {2016})}\BibitemShut
  {NoStop}%
\bibitem [{\citenamefont {Zyuzin}\ \emph
  {et~al.}(2016{\natexlab{b}})\citenamefont {Zyuzin}, \citenamefont
  {Alidoust},\ and\ \citenamefont {Loss}}]{Zyuzin2016}%
  \BibitemOpen
  \bibfield  {author} {\bibinfo {author} {\bibfnamefont {A.}~\bibnamefont
  {Zyuzin}}, \bibinfo {author} {\bibfnamefont {M.}~\bibnamefont {Alidoust}}, \
  and\ \bibinfo {author} {\bibfnamefont {D.}~\bibnamefont {Loss}},\ }\href
  {https://link.aps.org/doi/10.1103/PhysRevB.93.214502} {\bibfield  {journal}
  {\bibinfo  {journal} {Phys. Rev. B}\ }\textbf {\bibinfo {volume} {93}},\
  \bibinfo {pages} {214502} (\bibinfo {year} {2016}{\natexlab{b}})}\BibitemShut
  {NoStop}%
\bibitem [{\citenamefont {Silaev}\ \emph {et~al.}(2017)\citenamefont {Silaev},
  \citenamefont {Tokatly},\ and\ \citenamefont {Bergeret}}]{Silaev2017}%
  \BibitemOpen
  \bibfield  {author} {\bibinfo {author} {\bibfnamefont {M.~A.}\ \bibnamefont
  {Silaev}}, \bibinfo {author} {\bibfnamefont {I.~V.}\ \bibnamefont {Tokatly}},
  \ and\ \bibinfo {author} {\bibfnamefont {F.~S.}\ \bibnamefont {Bergeret}},\
  }\href {https://link.aps.org/doi/10.1103/PhysRevB.95.184508} {\bibfield
  {journal} {\bibinfo  {journal} {Phys. Rev. B}\ }\textbf {\bibinfo {volume}
  {95}},\ \bibinfo {pages} {184508} (\bibinfo {year} {2017})}\BibitemShut
  {NoStop}%
\bibitem [{\citenamefont {Bobkova}\ \emph {et~al.}(2017)\citenamefont
  {Bobkova}, \citenamefont {Bobkov},\ and\ \citenamefont
  {Silaev}}]{Bobkova2017}%
  \BibitemOpen
  \bibfield  {author} {\bibinfo {author} {\bibfnamefont {I.~V.}\ \bibnamefont
  {Bobkova}}, \bibinfo {author} {\bibfnamefont {A.~M.}\ \bibnamefont {Bobkov}},
  \ and\ \bibinfo {author} {\bibfnamefont {M.~A.}\ \bibnamefont {Silaev}},\
  }\href {https://link.aps.org/doi/10.1103/PhysRevB.96.094506} {\bibfield
  {journal} {\bibinfo  {journal} {Phys. Rev. B}\ }\textbf {\bibinfo {volume}
  {96}},\ \bibinfo {pages} {094506} (\bibinfo {year} {2017})}\BibitemShut
  {NoStop}%
\bibitem [{\citenamefont {Bobkova}\ \emph {et~al.}(2016)\citenamefont
  {Bobkova}, \citenamefont {Bobkov}, \citenamefont {Zyuzin},\ and\
  \citenamefont {Alidoust}}]{PhysRevB.94.134506}%
  \BibitemOpen
  \bibfield  {author} {\bibinfo {author} {\bibfnamefont {I.~V.}\ \bibnamefont
  {Bobkova}}, \bibinfo {author} {\bibfnamefont {A.~M.}\ \bibnamefont {Bobkov}},
  \bibinfo {author} {\bibfnamefont {A.~A.}\ \bibnamefont {Zyuzin}}, \ and\
  \bibinfo {author} {\bibfnamefont {M.}~\bibnamefont {Alidoust}},\ }\href
  {\doibase 10.1103/PhysRevB.94.134506} {\bibfield  {journal} {\bibinfo
  {journal} {Phys. Rev. B}\ }\textbf {\bibinfo {volume} {94}},\ \bibinfo
  {pages} {134506} (\bibinfo {year} {2016})}\BibitemShut {NoStop}%
\bibitem [{\citenamefont {Szombati}\ \emph {et~al.}(2016)\citenamefont
  {Szombati}, \citenamefont {Nadj-Perge}, \citenamefont {Car}, \citenamefont
  {Plissard}, \citenamefont {Bakkers},\ and\ \citenamefont
  {Kouwenhoven}}]{Szombati2016}%
  \BibitemOpen
  \bibfield  {author} {\bibinfo {author} {\bibfnamefont {D.~B.}\ \bibnamefont
  {Szombati}}, \bibinfo {author} {\bibfnamefont {S.}~\bibnamefont
  {Nadj-Perge}}, \bibinfo {author} {\bibfnamefont {D.}~\bibnamefont {Car}},
  \bibinfo {author} {\bibfnamefont {S.~R.}\ \bibnamefont {Plissard}}, \bibinfo
  {author} {\bibfnamefont {E.~P. A.~M.}\ \bibnamefont {Bakkers}}, \ and\
  \bibinfo {author} {\bibfnamefont {L.~P.}\ \bibnamefont {Kouwenhoven}},\
  }\href {http://dx.doi.org/10.1038/nphys3742} {\bibfield  {journal} {\bibinfo
  {journal} {Nature Physics}\ }\textbf {\bibinfo {volume} {12}},\ \bibinfo
  {pages} {568 EP } (\bibinfo {year} {2016})}\BibitemShut {NoStop}%
\bibitem [{\citenamefont {Assouline}\ \emph {et~al.}(2018)\citenamefont
  {Assouline}, \citenamefont {Feuillet-Palma}, \citenamefont {Bergeal},
  \citenamefont {Zhang}, \citenamefont {Mottaghizadeh}, \citenamefont
  {Zimmers}, \citenamefont {Lhuillier}, \citenamefont {Marangolo},
  \citenamefont {Eddrief}, \citenamefont {Atkinson}, \citenamefont {Aprili},\
  and\ \citenamefont {Aubin}}]{1806.01406}%
  \BibitemOpen
  \bibfield  {author} {\bibinfo {author} {\bibfnamefont {A.}~\bibnamefont
  {Assouline}}, \bibinfo {author} {\bibfnamefont {C.}~\bibnamefont
  {Feuillet-Palma}}, \bibinfo {author} {\bibfnamefont {N.}~\bibnamefont
  {Bergeal}}, \bibinfo {author} {\bibfnamefont {T.}~\bibnamefont {Zhang}},
  \bibinfo {author} {\bibfnamefont {A.}~\bibnamefont {Mottaghizadeh}}, \bibinfo
  {author} {\bibfnamefont {A.}~\bibnamefont {Zimmers}}, \bibinfo {author}
  {\bibfnamefont {E.}~\bibnamefont {Lhuillier}}, \bibinfo {author}
  {\bibfnamefont {M.}~\bibnamefont {Marangolo}}, \bibinfo {author}
  {\bibfnamefont {M.}~\bibnamefont {Eddrief}}, \bibinfo {author} {\bibfnamefont
  {P.}~\bibnamefont {Atkinson}}, \bibinfo {author} {\bibfnamefont
  {M.}~\bibnamefont {Aprili}}, \ and\ \bibinfo {author} {\bibfnamefont
  {H.}~\bibnamefont {Aubin}},\ }\href@noop {} {\enquote {\bibinfo {title}
  {Spin-orbit induced phase-shift in bi$_{2}$se$_{3}$ josephson junctions},}\ }
  (\bibinfo {year} {2018}),\ \Eprint {http://arxiv.org/abs/arXiv:1806.01406}
  {arXiv:1806.01406} \BibitemShut {NoStop}%
\bibitem [{\citenamefont {Murani}\ \emph {et~al.}(2017)\citenamefont {Murani},
  \citenamefont {Kasumov}, \citenamefont {Sengupta}, \citenamefont {Kasumov},
  \citenamefont {Volkov}, \citenamefont {Khodos}, \citenamefont {Brisset},
  \citenamefont {Delagrange}, \citenamefont {Chepelianskii}, \citenamefont
  {Deblock}, \citenamefont {Bouchiat},\ and\ \citenamefont
  {Guéron}}]{Murani2017}%
  \BibitemOpen
  \bibfield  {author} {\bibinfo {author} {\bibfnamefont {A.}~\bibnamefont
  {Murani}}, \bibinfo {author} {\bibfnamefont {A.}~\bibnamefont {Kasumov}},
  \bibinfo {author} {\bibfnamefont {S.}~\bibnamefont {Sengupta}}, \bibinfo
  {author} {\bibfnamefont {Y.~A.}\ \bibnamefont {Kasumov}}, \bibinfo {author}
  {\bibfnamefont {V.~T.}\ \bibnamefont {Volkov}}, \bibinfo {author}
  {\bibfnamefont {I.~I.}\ \bibnamefont {Khodos}}, \bibinfo {author}
  {\bibfnamefont {F.}~\bibnamefont {Brisset}}, \bibinfo {author} {\bibfnamefont
  {R.}~\bibnamefont {Delagrange}}, \bibinfo {author} {\bibfnamefont
  {A.}~\bibnamefont {Chepelianskii}}, \bibinfo {author} {\bibfnamefont
  {R.}~\bibnamefont {Deblock}}, \bibinfo {author} {\bibfnamefont
  {H.}~\bibnamefont {Bouchiat}}, \ and\ \bibinfo {author} {\bibfnamefont
  {S.}~\bibnamefont {Guéron}},\ }\href {http://dx.doi.org/10.1038/ncomms15941}
  {\bibfield  {journal} {\bibinfo  {journal} {Nature Communications}\ }\textbf
  {\bibinfo {volume} {8}},\ \bibinfo {pages} {15941} (\bibinfo {year}
  {2017})}\BibitemShut {NoStop}%
\bibitem [{\citenamefont {Glick}\ \emph {et~al.}(2018)\citenamefont {Glick},
  \citenamefont {Aguilar}, \citenamefont {Gougam}, \citenamefont {Niedzielski},
  \citenamefont {Gingrich}, \citenamefont {Loloee}, \citenamefont {Pratt},\
  and\ \citenamefont {Birge}}]{Glickeaat9457}%
  \BibitemOpen
  \bibfield  {author} {\bibinfo {author} {\bibfnamefont {J.~A.}\ \bibnamefont
  {Glick}}, \bibinfo {author} {\bibfnamefont {V.}~\bibnamefont {Aguilar}},
  \bibinfo {author} {\bibfnamefont {A.~B.}\ \bibnamefont {Gougam}}, \bibinfo
  {author} {\bibfnamefont {B.~M.}\ \bibnamefont {Niedzielski}}, \bibinfo
  {author} {\bibfnamefont {E.~C.}\ \bibnamefont {Gingrich}}, \bibinfo {author}
  {\bibfnamefont {R.}~\bibnamefont {Loloee}}, \bibinfo {author} {\bibfnamefont
  {W.~P.}\ \bibnamefont {Pratt}}, \ and\ \bibinfo {author} {\bibfnamefont
  {N.~O.}\ \bibnamefont {Birge}},\ }\href {\doibase 10.1126/sciadv.aat9457}
  {\bibfield  {journal} {\bibinfo  {journal} {Science Advances}\ }\textbf
  {\bibinfo {volume} {4}} (\bibinfo {year} {2018}),\ 10.1126/sciadv.aat9457},\
  \Eprint
  {http://arxiv.org/abs/http://advances.sciencemag.org/content/4/7/eaat9457.full.pdf}
  {http://advances.sciencemag.org/content/4/7/eaat9457.full.pdf} \BibitemShut
  {NoStop}%
\bibitem [{\citenamefont {Bergeret}\ \emph
  {et~al.}(2005{\natexlab{a}})\citenamefont {Bergeret}, \citenamefont
  {Yeyati},\ and\ \citenamefont {Martín-Rodero}}]{Bergeret2005a}%
  \BibitemOpen
  \bibfield  {author} {\bibinfo {author} {\bibfnamefont {F.~S.}\ \bibnamefont
  {Bergeret}}, \bibinfo {author} {\bibfnamefont {A.~L.}\ \bibnamefont
  {Yeyati}}, \ and\ \bibinfo {author} {\bibfnamefont {A.}~\bibnamefont
  {Martín-Rodero}},\ }\href
  {https://link.aps.org/doi/10.1103/PhysRevB.72.064524} {\bibfield  {journal}
  {\bibinfo  {journal} {Phys. Rev. B}\ }\textbf {\bibinfo {volume} {72}},\
  \bibinfo {pages} {064524} (\bibinfo {year} {2005}{\natexlab{a}})}\BibitemShut
  {NoStop}%
\bibitem [{\citenamefont {Lee}\ \emph {et~al.}(2007)\citenamefont {Lee},
  \citenamefont {Onose}, \citenamefont {Tokura},\ and\ \citenamefont
  {Ong}}]{Lee2007}%
  \BibitemOpen
  \bibfield  {author} {\bibinfo {author} {\bibfnamefont {M.}~\bibnamefont
  {Lee}}, \bibinfo {author} {\bibfnamefont {Y.}~\bibnamefont {Onose}}, \bibinfo
  {author} {\bibfnamefont {Y.}~\bibnamefont {Tokura}}, \ and\ \bibinfo {author}
  {\bibfnamefont {N.~P.}\ \bibnamefont {Ong}},\ }\href {\doibase
  https://doi.org/10.1103/PhysRevB.75.172403} {\bibfield  {journal} {\bibinfo
  {journal} {Phys. Rev. B}\ }\textbf {\bibinfo {volume} {75}},\ \bibinfo
  {pages} {172403} (\bibinfo {year} {2007})}\BibitemShut {NoStop}%
\bibitem [{\citenamefont {Jonietz}\ \emph {et~al.}(2010)\citenamefont
  {Jonietz}, \citenamefont {M${\rm \ddot u}$hlbauer}, \citenamefont
  {Pfleiderer}, \citenamefont {Neubauer}, \citenamefont {M${\rm \ddot u}$nzer},
  \citenamefont {Bauer}, \citenamefont {Adams}, \citenamefont {Georgii},
  \citenamefont {Böni}, \citenamefont {Duine}, \citenamefont {Everschor},
  \citenamefont {Garst},\ and\ \citenamefont {Rosch}}]{Jonietz2010}%
  \BibitemOpen
  \bibfield  {author} {\bibinfo {author} {\bibfnamefont {F.}~\bibnamefont
  {Jonietz}}, \bibinfo {author} {\bibfnamefont {S.}~\bibnamefont {M${\rm \ddot
  u}$hlbauer}}, \bibinfo {author} {\bibfnamefont {C.}~\bibnamefont
  {Pfleiderer}}, \bibinfo {author} {\bibfnamefont {A.}~\bibnamefont
  {Neubauer}}, \bibinfo {author} {\bibfnamefont {W.}~\bibnamefont {M${\rm \ddot
  u}$nzer}}, \bibinfo {author} {\bibfnamefont {A.}~\bibnamefont {Bauer}},
  \bibinfo {author} {\bibfnamefont {T.}~\bibnamefont {Adams}}, \bibinfo
  {author} {\bibfnamefont {R.}~\bibnamefont {Georgii}}, \bibinfo {author}
  {\bibfnamefont {P.}~\bibnamefont {Böni}}, \bibinfo {author} {\bibfnamefont
  {R.~A.}\ \bibnamefont {Duine}}, \bibinfo {author} {\bibfnamefont
  {K.}~\bibnamefont {Everschor}}, \bibinfo {author} {\bibfnamefont
  {M.}~\bibnamefont {Garst}}, \ and\ \bibinfo {author} {\bibfnamefont
  {A.}~\bibnamefont {Rosch}},\ }\href
  {http://science.sciencemag.org/content/330/6011/1648.abstract} {\bibfield
  {journal} {\bibinfo  {journal} {Science}\ }\textbf {\bibinfo {volume}
  {330}},\ \bibinfo {pages} {1648} (\bibinfo {year} {2010})}\BibitemShut
  {NoStop}%
\bibitem [{\citenamefont {Volkov}\ \emph {et~al.}(2006)\citenamefont {Volkov},
  \citenamefont {Anishchanka},\ and\ \citenamefont {Efetov}}]{Volkov2006}%
  \BibitemOpen
  \bibfield  {author} {\bibinfo {author} {\bibfnamefont {A.~F.}\ \bibnamefont
  {Volkov}}, \bibinfo {author} {\bibfnamefont {A.}~\bibnamefont {Anishchanka}},
  \ and\ \bibinfo {author} {\bibfnamefont {K.~B.}\ \bibnamefont {Efetov}},\
  }\href {http://link.aps.org/doi/10.1103/PhysRevB.73.104412} {\bibfield
  {journal} {\bibinfo  {journal} {Phys. Rev. B}\ }\textbf {\bibinfo {volume}
  {73}},\ \bibinfo {pages} {104412} (\bibinfo {year} {2006})}\BibitemShut
  {NoStop}%
\bibitem [{\citenamefont {Kalenkov}\ \emph {et~al.}(2011)\citenamefont
  {Kalenkov}, \citenamefont {Zaikin},\ and\ \citenamefont
  {Petrashov}}]{Kalenkov2011}%
  \BibitemOpen
  \bibfield  {author} {\bibinfo {author} {\bibfnamefont {M.~S.}\ \bibnamefont
  {Kalenkov}}, \bibinfo {author} {\bibfnamefont {A.~D.}\ \bibnamefont
  {Zaikin}}, \ and\ \bibinfo {author} {\bibfnamefont {V.~T.}\ \bibnamefont
  {Petrashov}},\ }\href
  {http://link.aps.org/doi/10.1103/PhysRevLett.107.087003} {\bibfield
  {journal} {\bibinfo  {journal} {Phys. Rev. Lett.}\ }\textbf {\bibinfo
  {volume} {107}},\ \bibinfo {pages} {087003} (\bibinfo {year}
  {2011})}\BibitemShut {NoStop}%
\bibitem [{\citenamefont {Yokoyama}\ and\ \citenamefont
  {Linder}(2015)}]{Yokoyama2015}%
  \BibitemOpen
  \bibfield  {author} {\bibinfo {author} {\bibfnamefont {T.}~\bibnamefont
  {Yokoyama}}\ and\ \bibinfo {author} {\bibfnamefont {J.}~\bibnamefont
  {Linder}},\ }\href {http://link.aps.org/doi/10.1103/PhysRevB.92.060503}
  {\bibfield  {journal} {\bibinfo  {journal} {Phys. Rev. B}\ }\textbf {\bibinfo
  {volume} {92}},\ \bibinfo {pages} {060503} (\bibinfo {year}
  {2015})}\BibitemShut {NoStop}%
\bibitem [{\citenamefont {Buzdin}(2005)}]{Buzdin2005}%
  \BibitemOpen
  \bibfield  {author} {\bibinfo {author} {\bibfnamefont {A.~I.}\ \bibnamefont
  {Buzdin}},\ }\href {https://link.aps.org/doi/10.1103/RevModPhys.77.935}
  {\bibfield  {journal} {\bibinfo  {journal} {Rev. Mod. Phys.}\ }\textbf
  {\bibinfo {volume} {77}},\ \bibinfo {pages} {935} (\bibinfo {year}
  {2005})}\BibitemShut {NoStop}%
\bibitem [{\citenamefont {Bergeret}\ \emph
  {et~al.}(2005{\natexlab{b}})\citenamefont {Bergeret}, \citenamefont
  {Volkov},\ and\ \citenamefont {Efetov}}]{bergeret2005odd}%
  \BibitemOpen
  \bibfield  {author} {\bibinfo {author} {\bibfnamefont {F.}~\bibnamefont
  {Bergeret}}, \bibinfo {author} {\bibfnamefont {A.}~\bibnamefont {Volkov}}, \
  and\ \bibinfo {author} {\bibfnamefont {K.}~\bibnamefont {Efetov}},\
  }\href@noop {} {\bibfield  {journal} {\bibinfo  {journal} {Rev. Mod. Phys.}\
  }\textbf {\bibinfo {volume} {77}},\ \bibinfo {pages} {1321} (\bibinfo {year}
  {2005}{\natexlab{b}})}\BibitemShut {NoStop}%
\bibitem [{\citenamefont {Bergeret}\ \emph
  {et~al.}(2012{\natexlab{a}})\citenamefont {Bergeret}, \citenamefont {Verso},\
  and\ \citenamefont {Volkov}}]{bergeret2012electronic}%
  \BibitemOpen
  \bibfield  {author} {\bibinfo {author} {\bibfnamefont {F.}~\bibnamefont
  {Bergeret}}, \bibinfo {author} {\bibfnamefont {A.}~\bibnamefont {Verso}}, \
  and\ \bibinfo {author} {\bibfnamefont {A.~F.}\ \bibnamefont {Volkov}},\
  }\href@noop {} {\bibfield  {journal} {\bibinfo  {journal} {Phys. Rev. B}\
  }\textbf {\bibinfo {volume} {86}},\ \bibinfo {pages} {214516} (\bibinfo
  {year} {2012}{\natexlab{a}})}\BibitemShut {NoStop}%
\bibitem [{\citenamefont {Bergeret}\ \emph
  {et~al.}(2012{\natexlab{b}})\citenamefont {Bergeret}, \citenamefont {Verso},\
  and\ \citenamefont {Volkov}}]{Bergeret2012a}%
  \BibitemOpen
  \bibfield  {author} {\bibinfo {author} {\bibfnamefont {F.~S.}\ \bibnamefont
  {Bergeret}}, \bibinfo {author} {\bibfnamefont {A.}~\bibnamefont {Verso}}, \
  and\ \bibinfo {author} {\bibfnamefont {A.~F.}\ \bibnamefont {Volkov}},\
  }\href {https://link.aps.org/doi/10.1103/PhysRevB.86.060506} {\bibfield
  {journal} {\bibinfo  {journal} {Phys. Rev. B}\ }\textbf {\bibinfo {volume}
  {86}},\ \bibinfo {pages} {060506} (\bibinfo {year}
  {2012}{\natexlab{b}})}\BibitemShut {NoStop}%
\bibitem [{\citenamefont {Eschrig}\ \emph
  {et~al.}(2015{\natexlab{a}})\citenamefont {Eschrig}, \citenamefont {Cottet},
  \citenamefont {Belzig},\ and\ \citenamefont {Linder}}]{Eschrig2015}%
  \BibitemOpen
  \bibfield  {author} {\bibinfo {author} {\bibfnamefont {M.}~\bibnamefont
  {Eschrig}}, \bibinfo {author} {\bibfnamefont {A.}~\bibnamefont {Cottet}},
  \bibinfo {author} {\bibfnamefont {W.}~\bibnamefont {Belzig}}, \ and\ \bibinfo
  {author} {\bibfnamefont {J.}~\bibnamefont {Linder}},\ }\href
  {http://stacks.iop.org/1367-2630/17/i=8/a=083037} {\bibfield  {journal}
  {\bibinfo  {journal} {New Journal of Physics}\ }\textbf {\bibinfo {volume}
  {17}},\ \bibinfo {pages} {083037} (\bibinfo {year}
  {2015}{\natexlab{a}})}\BibitemShut {NoStop}%
\bibitem [{\citenamefont {Ishikawa}\ \emph {et~al.}(1977)\citenamefont
  {Ishikawa}, \citenamefont {Shirane}, \citenamefont {Tarvin},\ and\
  \citenamefont {Kohgi}}]{Ishikawa1977}%
  \BibitemOpen
  \bibfield  {author} {\bibinfo {author} {\bibfnamefont {Y.}~\bibnamefont
  {Ishikawa}}, \bibinfo {author} {\bibfnamefont {G.}~\bibnamefont {Shirane}},
  \bibinfo {author} {\bibfnamefont {J.~A.}\ \bibnamefont {Tarvin}}, \ and\
  \bibinfo {author} {\bibfnamefont {M.}~\bibnamefont {Kohgi}},\ }\href
  {https://link.aps.org/doi/10.1103/PhysRevB.16.4956} {\bibfield  {journal}
  {\bibinfo  {journal} {Phys. Rev. B}\ }\textbf {\bibinfo {volume} {16}},\
  \bibinfo {pages} {4956} (\bibinfo {year} {1977})}\BibitemShut {NoStop}%
\bibitem [{\citenamefont {Pfleiderer}\ \emph {et~al.}(2001)\citenamefont
  {Pfleiderer}, \citenamefont {Julian},\ and\ \citenamefont
  {Lonzarich}}]{Pfleiderer2001}%
  \BibitemOpen
  \bibfield  {author} {\bibinfo {author} {\bibfnamefont {C.}~\bibnamefont
  {Pfleiderer}}, \bibinfo {author} {\bibfnamefont {S.~R.}\ \bibnamefont
  {Julian}}, \ and\ \bibinfo {author} {\bibfnamefont {G.~G.}\ \bibnamefont
  {Lonzarich}},\ }\href {http://dx.doi.org/10.1038/35106527} {\bibfield
  {journal} {\bibinfo  {journal} {Nature}\ }\textbf {\bibinfo {volume} {414}},\
  \bibinfo {pages} {427} (\bibinfo {year} {2001})}\BibitemShut {NoStop}%
\bibitem [{\citenamefont {Uchida}\ \emph {et~al.}(2006)\citenamefont {Uchida},
  \citenamefont {Onose}, \citenamefont {Matsui},\ and\ \citenamefont
  {Tokura}}]{Uchida2006}%
  \BibitemOpen
  \bibfield  {author} {\bibinfo {author} {\bibfnamefont {M.}~\bibnamefont
  {Uchida}}, \bibinfo {author} {\bibfnamefont {Y.}~\bibnamefont {Onose}},
  \bibinfo {author} {\bibfnamefont {Y.}~\bibnamefont {Matsui}}, \ and\ \bibinfo
  {author} {\bibfnamefont {Y.}~\bibnamefont {Tokura}},\ }\href
  {http://science.sciencemag.org/content/311/5759/359.abstract} {\bibfield
  {journal} {\bibinfo  {journal} {Science}\ }\textbf {\bibinfo {volume}
  {311}},\ \bibinfo {pages} {359} (\bibinfo {year} {2006})}\BibitemShut
  {NoStop}%
\bibitem [{\citenamefont {Kuprianov}\ and\ \citenamefont
  {Lukichev}(1988)}]{KL}%
  \BibitemOpen
  \bibfield  {author} {\bibinfo {author} {\bibfnamefont {M.~Y.}\ \bibnamefont
  {Kuprianov}}\ and\ \bibinfo {author} {\bibfnamefont {V.~F.}\ \bibnamefont
  {Lukichev}},\ }\href@noop {} {\bibfield  {journal} {\bibinfo  {journal}
  {JETP}\ }\textbf {\bibinfo {volume} {67}},\ \bibinfo {pages} {1163} (\bibinfo
  {year} {1988})},\ \bibinfo {note} {[Zh. Eksp. Teor. Fiz. {\bf 94}, 139
  (1988)]}\BibitemShut {NoStop}%
\bibitem [{\citenamefont {Bergeret}\ \emph
  {et~al.}(2012{\natexlab{c}})\citenamefont {Bergeret}, \citenamefont {Verso},\
  and\ \citenamefont {Volkov}}]{BergeretVerso2012}%
  \BibitemOpen
  \bibfield  {author} {\bibinfo {author} {\bibfnamefont {F.~S.}\ \bibnamefont
  {Bergeret}}, \bibinfo {author} {\bibfnamefont {A.}~\bibnamefont {Verso}}, \
  and\ \bibinfo {author} {\bibfnamefont {A.~F.}\ \bibnamefont {Volkov}},\
  }\href {\doibase https://doi.org/10.1103/PhysRevB.86.214516} {\bibfield
  {journal} {\bibinfo  {journal} {Phys. Rev. B}\ }\textbf {\bibinfo {volume}
  {86}},\ \bibinfo {pages} {214516} (\bibinfo {year}
  {2012}{\natexlab{c}})}\BibitemShut {NoStop}%
\bibitem [{\citenamefont {Machon}\ \emph {et~al.}(2013)\citenamefont {Machon},
  \citenamefont {Eschrig},\ and\ \citenamefont {Belzig}}]{Machon2013}%
  \BibitemOpen
  \bibfield  {author} {\bibinfo {author} {\bibfnamefont {P.}~\bibnamefont
  {Machon}}, \bibinfo {author} {\bibfnamefont {M.}~\bibnamefont {Eschrig}}, \
  and\ \bibinfo {author} {\bibfnamefont {W.}~\bibnamefont {Belzig}},\ }\href
  {\doibase 10.1103/PhysRevLett.110.047002} {\bibfield  {journal} {\bibinfo
  {journal} {Phys. Rev. Lett.}\ }\textbf {\bibinfo {volume} {110}},\ \bibinfo
  {pages} {047002} (\bibinfo {year} {2013})}\BibitemShut {NoStop}%
\bibitem [{\citenamefont {Machon}\ \emph {et~al.}(2014)\citenamefont {Machon},
  \citenamefont {M.},\ and\ \citenamefont {Belzig}}]{Machon2014}%
  \BibitemOpen
  \bibfield  {author} {\bibinfo {author} {\bibfnamefont {P.}~\bibnamefont
  {Machon}}, \bibinfo {author} {\bibfnamefont {E.}~\bibnamefont {M.}}, \ and\
  \bibinfo {author} {\bibfnamefont {W.}~\bibnamefont {Belzig}},\ }\href
  {\doibase https://doi.org/10.1088/1367-2630/16/7/073002} {\bibfield
  {journal} {\bibinfo  {journal} {New J. Phys.}\ }\textbf {\bibinfo {volume}
  {16}},\ \bibinfo {pages} {073002} (\bibinfo {year} {2014})}\BibitemShut
  {NoStop}%
\bibitem [{\citenamefont {Eschrig}\ \emph
  {et~al.}(2015{\natexlab{b}})\citenamefont {Eschrig}, \citenamefont {Cottet},
  \citenamefont {Belzig},\ and\ \citenamefont {Linder}}]{EschrigBC2015}%
  \BibitemOpen
  \bibfield  {author} {\bibinfo {author} {\bibfnamefont {M.}~\bibnamefont
  {Eschrig}}, \bibinfo {author} {\bibfnamefont {A.}~\bibnamefont {Cottet}},
  \bibinfo {author} {\bibfnamefont {W.}~\bibnamefont {Belzig}}, \ and\ \bibinfo
  {author} {\bibfnamefont {J.}~\bibnamefont {Linder}},\ }\href {\doibase
  https://doi.org/10.1088/1367-2630/17/8/083037} {\bibfield  {journal}
  {\bibinfo  {journal} {New J. Phys.}\ }\textbf {\bibinfo {volume} {17}},\
  \bibinfo {pages} {083037} (\bibinfo {year} {2015}{\natexlab{b}})}\BibitemShut
  {NoStop}%
\bibitem [{\citenamefont {Silaev}(2017)}]{PhysRevB.96.064519}%
  \BibitemOpen
  \bibfield  {author} {\bibinfo {author} {\bibfnamefont {M.~A.}\ \bibnamefont
  {Silaev}},\ }\href {\doibase 10.1103/PhysRevB.96.064519} {\bibfield
  {journal} {\bibinfo  {journal} {Phys. Rev. B}\ }\textbf {\bibinfo {volume}
  {96}},\ \bibinfo {pages} {064519} (\bibinfo {year} {2017})}\BibitemShut
  {NoStop}%
\bibitem [{\citenamefont {Nagaosa}\ and\ \citenamefont
  {Tokura}(2013{\natexlab{b}})}]{Nagaosa-NatureNano2013}%
  \BibitemOpen
  \bibfield  {author} {\bibinfo {author} {\bibfnamefont {N.}~\bibnamefont
  {Nagaosa}}\ and\ \bibinfo {author} {\bibfnamefont {Y.}~\bibnamefont
  {Tokura}},\ }\href {http://dx.doi.org/10.1038/nnano.2013.243} {\bibfield
  {journal} {\bibinfo  {journal} {Nat Nano}\ }\textbf {\bibinfo {volume} {8}},\
  \bibinfo {pages} {899} (\bibinfo {year} {2013}{\natexlab{b}})}\BibitemShut
  {NoStop}%
\bibitem [{\citenamefont {Roszler}\ \emph {et~al.}(2006)\citenamefont
  {Roszler}, \citenamefont {Bogdanov},\ and\ \citenamefont
  {Pfleiderer}}]{Roszler2006}%
  \BibitemOpen
  \bibfield  {author} {\bibinfo {author} {\bibfnamefont {U.~K.}\ \bibnamefont
  {Roszler}}, \bibinfo {author} {\bibfnamefont {A.~N.}\ \bibnamefont
  {Bogdanov}}, \ and\ \bibinfo {author} {\bibfnamefont {C.}~\bibnamefont
  {Pfleiderer}},\ }\href {http://dx.doi.org/10.1038/nature05056} {\bibfield
  {journal} {\bibinfo  {journal} {Nature}\ }\textbf {\bibinfo {volume} {442}},\
  \bibinfo {pages} {797} (\bibinfo {year} {2006})}\BibitemShut {NoStop}%
\bibitem [{\citenamefont {M${\rm \ddot u}$hlbauer}\ \emph
  {et~al.}(2009)\citenamefont {M${\rm \ddot u}$hlbauer}, \citenamefont {Binz},
  \citenamefont {Jonietz}, \citenamefont {Pfleiderer}, \citenamefont {Rosch},
  \citenamefont {Neubauer}, \citenamefont {Georgii},\ and\ \citenamefont
  {B${\rm \ddot o}$ni}}]{Muehlbauer2009}%
  \BibitemOpen
  \bibfield  {author} {\bibinfo {author} {\bibfnamefont {S.}~\bibnamefont
  {M${\rm \ddot u}$hlbauer}}, \bibinfo {author} {\bibfnamefont
  {B.}~\bibnamefont {Binz}}, \bibinfo {author} {\bibfnamefont {F.}~\bibnamefont
  {Jonietz}}, \bibinfo {author} {\bibfnamefont {C.}~\bibnamefont {Pfleiderer}},
  \bibinfo {author} {\bibfnamefont {A.}~\bibnamefont {Rosch}}, \bibinfo
  {author} {\bibfnamefont {A.}~\bibnamefont {Neubauer}}, \bibinfo {author}
  {\bibfnamefont {R.}~\bibnamefont {Georgii}}, \ and\ \bibinfo {author}
  {\bibfnamefont {P.}~\bibnamefont {B${\rm \ddot o}$ni}},\ }\href
  {http://science.sciencemag.org/content/323/5916/915.abstract} {\bibfield
  {journal} {\bibinfo  {journal} {Science}\ }\textbf {\bibinfo {volume}
  {323}},\ \bibinfo {pages} {915} (\bibinfo {year} {2009})}\BibitemShut
  {NoStop}%
\bibitem [{\citenamefont {Yu}\ \emph {et~al.}(2010)\citenamefont {Yu},
  \citenamefont {Onose}, \citenamefont {Kanazawa}, \citenamefont {Park},
  \citenamefont {Han}, \citenamefont {Matsui}, \citenamefont {Nagaosa},\ and\
  \citenamefont {Tokura}}]{Yu2010}%
  \BibitemOpen
  \bibfield  {author} {\bibinfo {author} {\bibfnamefont {X.~Z.}\ \bibnamefont
  {Yu}}, \bibinfo {author} {\bibfnamefont {Y.}~\bibnamefont {Onose}}, \bibinfo
  {author} {\bibfnamefont {N.}~\bibnamefont {Kanazawa}}, \bibinfo {author}
  {\bibfnamefont {J.~H.}\ \bibnamefont {Park}}, \bibinfo {author}
  {\bibfnamefont {J.~H.}\ \bibnamefont {Han}}, \bibinfo {author} {\bibfnamefont
  {Y.}~\bibnamefont {Matsui}}, \bibinfo {author} {\bibfnamefont
  {N.}~\bibnamefont {Nagaosa}}, \ and\ \bibinfo {author} {\bibfnamefont
  {Y.}~\bibnamefont {Tokura}},\ }\href {http://dx.doi.org/10.1038/nature09124}
  {\bibfield  {journal} {\bibinfo  {journal} {Nature}\ }\textbf {\bibinfo
  {volume} {465}},\ \bibinfo {pages} {901} (\bibinfo {year}
  {2010})}\BibitemShut {NoStop}%
\bibitem [{\citenamefont {Parkin}\ \emph {et~al.}(2008)\citenamefont {Parkin},
  \citenamefont {Hayashi},\ and\ \citenamefont {Thomas}}]{Parkin2008}%
  \BibitemOpen
  \bibfield  {author} {\bibinfo {author} {\bibfnamefont {S.~S.~P.}\
  \bibnamefont {Parkin}}, \bibinfo {author} {\bibfnamefont {M.}~\bibnamefont
  {Hayashi}}, \ and\ \bibinfo {author} {\bibfnamefont {L.}~\bibnamefont
  {Thomas}},\ }\href
  {http://science.sciencemag.org/content/320/5873/190.abstract} {\bibfield
  {journal} {\bibinfo  {journal} {Science}\ }\textbf {\bibinfo {volume}
  {320}},\ \bibinfo {pages} {190} (\bibinfo {year} {2008})}\BibitemShut
  {NoStop}%
\bibitem [{\citenamefont {Yu}\ \emph {et~al.}(2012)\citenamefont {Yu},
  \citenamefont {Kanazawa}, \citenamefont {Zhang}, \citenamefont {Nagai},
  \citenamefont {Hara}, \citenamefont {Kimoto}, \citenamefont {Matsui},
  \citenamefont {Onose},\ and\ \citenamefont {Tokura}}]{Yu2012}%
  \BibitemOpen
  \bibfield  {author} {\bibinfo {author} {\bibfnamefont {X.~Z.}\ \bibnamefont
  {Yu}}, \bibinfo {author} {\bibfnamefont {N.}~\bibnamefont {Kanazawa}},
  \bibinfo {author} {\bibfnamefont {W.~Z.}\ \bibnamefont {Zhang}}, \bibinfo
  {author} {\bibfnamefont {T.}~\bibnamefont {Nagai}}, \bibinfo {author}
  {\bibfnamefont {T.}~\bibnamefont {Hara}}, \bibinfo {author} {\bibfnamefont
  {K.}~\bibnamefont {Kimoto}}, \bibinfo {author} {\bibfnamefont
  {Y.}~\bibnamefont {Matsui}}, \bibinfo {author} {\bibfnamefont
  {Y.}~\bibnamefont {Onose}}, \ and\ \bibinfo {author} {\bibfnamefont
  {Y.}~\bibnamefont {Tokura}},\ }\href {http://dx.doi.org/10.1038/ncomms1990}
  {\bibfield  {journal} {\bibinfo  {journal} {Nature Communications}\ }\textbf
  {\bibinfo {volume} {3}},\ \bibinfo {pages} {988} (\bibinfo {year}
  {2012})}\BibitemShut {NoStop}%
\bibitem [{\citenamefont {Hecht}(2012)}]{Hecht2012}%
  \BibitemOpen
  \bibfield  {author} {\bibinfo {author} {\bibfnamefont {F.}~\bibnamefont
  {Hecht}},\ }\href@noop {} {\bibfield  {journal} {\bibinfo  {journal} {J.
  Numer. Math.}\ }\textbf {\bibinfo {volume} {20}},\ \bibinfo {pages} {251}
  (\bibinfo {year} {2012})}\BibitemShut {NoStop}%
\bibitem [{\citenamefont {Bobkova}\ \emph {et~al.}(2018)\citenamefont
  {Bobkova}, \citenamefont {Bobkov},\ and\ \citenamefont
  {Silaev}}]{PhysRevB.98.014521}%
  \BibitemOpen
  \bibfield  {author} {\bibinfo {author} {\bibfnamefont {I.~V.}\ \bibnamefont
  {Bobkova}}, \bibinfo {author} {\bibfnamefont {A.~M.}\ \bibnamefont {Bobkov}},
  \ and\ \bibinfo {author} {\bibfnamefont {M.~A.}\ \bibnamefont {Silaev}},\
  }\href {\doibase 10.1103/PhysRevB.98.014521} {\bibfield  {journal} {\bibinfo
  {journal} {Phys. Rev. B}\ }\textbf {\bibinfo {volume} {98}},\ \bibinfo
  {pages} {014521} (\bibinfo {year} {2018})}\BibitemShut {NoStop}%
\bibitem [{\citenamefont {Heide}\ \emph {et~al.}(2008)\citenamefont {Heide},
  \citenamefont {Bihlmayer},\ and\ \citenamefont {Blugel}}]{Heide2008}%
  \BibitemOpen
  \bibfield  {author} {\bibinfo {author} {\bibfnamefont {M.}~\bibnamefont
  {Heide}}, \bibinfo {author} {\bibfnamefont {G.}~\bibnamefont {Bihlmayer}}, \
  and\ \bibinfo {author} {\bibfnamefont {S.}~\bibnamefont {Blugel}},\ }\href
  {https://doi.org/10.1103/PhysRevB.78.140403} {\bibfield  {journal} {\bibinfo
  {journal} {Phys. Rev. B}\ }\textbf {\bibinfo {volume} {78}},\ \bibinfo
  {pages} {140403} (\bibinfo {year} {2008})}\BibitemShut {NoStop}%
\bibitem [{\citenamefont {Thiaville}\ \emph {et~al.}(2012)\citenamefont
  {Thiaville}, \citenamefont {Rohart}, \citenamefont {Jue}, \citenamefont
  {Cros},\ and\ \citenamefont {Fert}}]{Thiaville2012}%
  \BibitemOpen
  \bibfield  {author} {\bibinfo {author} {\bibfnamefont {A.}~\bibnamefont
  {Thiaville}}, \bibinfo {author} {\bibfnamefont {S.}~\bibnamefont {Rohart}},
  \bibinfo {author} {\bibfnamefont {E.}~\bibnamefont {Jue}}, \bibinfo {author}
  {\bibfnamefont {V.}~\bibnamefont {Cros}}, \ and\ \bibinfo {author}
  {\bibfnamefont {A.}~\bibnamefont {Fert}},\ }\href
  {https://doi.org/10.1209/0295-5075/100/57002} {\bibfield  {journal} {\bibinfo
   {journal} {Europhys. Lett.}\ }\textbf {\bibinfo {volume} {100}},\ \bibinfo
  {pages} {57002} (\bibinfo {year} {2012})}\BibitemShut {NoStop}%
\bibitem [{\citenamefont {Emori}\ \emph {et~al.}(2013)\citenamefont {Emori},
  \citenamefont {Bauer}, \citenamefont {Ahn}, \citenamefont {Martinez},\ and\
  \citenamefont {D.}}]{Emori2013}%
  \BibitemOpen
  \bibfield  {author} {\bibinfo {author} {\bibfnamefont {S.}~\bibnamefont
  {Emori}}, \bibinfo {author} {\bibfnamefont {U.}~\bibnamefont {Bauer}},
  \bibinfo {author} {\bibfnamefont {S.-M.}\ \bibnamefont {Ahn}}, \bibinfo
  {author} {\bibfnamefont {E.}~\bibnamefont {Martinez}}, \ and\ \bibinfo
  {author} {\bibfnamefont {B.~G.~S.}\ \bibnamefont {D.}},\ }\href
  {https://doi.org/10.1038/nmat3675} {\bibfield  {journal} {\bibinfo  {journal}
  {Nat. Mater.}\ }\textbf {\bibinfo {volume} {12}},\ \bibinfo {pages} {611}
  (\bibinfo {year} {2013})}\BibitemShut {NoStop}%
\end{thebibliography}%

 \end{document}